\pdfoutput=1
\documentclass[12pt,a4paper]{article}

\usepackage{ifthen} 
\newboolean{pdflatex}
\setboolean{pdflatex}{true} 
\newboolean{articletitles}
\setboolean{articletitles}{true} 
\newboolean{uprightparticles}
\setboolean{uprightparticles}{false} 
\newboolean{inbibliography}
\setboolean{inbibliography}{false} 

\usepackage{microtype}
\usepackage{lineno}  
\usepackage{xspace} 
\usepackage{caption} 

\usepackage{graphicx}  
\usepackage{color}
\usepackage{colortbl}
\graphicspath{{./figs/}} 

\usepackage{amsmath} 
\usepackage{amssymb}
\usepackage{amsfonts}
\usepackage{upgreek} 

\newcommand*\patchAmsMathEnvironmentForLineno[1]{%
\expandafter\let\csname old#1\expandafter\endcsname\csname #1\endcsname
\expandafter\let\csname oldend#1\expandafter\endcsname\csname
end#1\endcsname
 \renewenvironment{#1}%
   {\linenomath\csname old#1\endcsname}%
   {\csname oldend#1\endcsname\endlinenomath}%
}
\newcommand*\patchBothAmsMathEnvironmentsForLineno[1]{%
  \patchAmsMathEnvironmentForLineno{#1}%
  \patchAmsMathEnvironmentForLineno{#1*}%
}
\AtBeginDocument{%
\patchBothAmsMathEnvironmentsForLineno{equation}%
\patchBothAmsMathEnvironmentsForLineno{align}%
\patchBothAmsMathEnvironmentsForLineno{flalign}%
\patchBothAmsMathEnvironmentsForLineno{alignat}%
\patchBothAmsMathEnvironmentsForLineno{gather}%
\patchBothAmsMathEnvironmentsForLineno{multline}%
\patchBothAmsMathEnvironmentsForLineno{eqnarray}%
}

\usepackage{hyperref}    
\usepackage[all]{hypcap} 

\def\lhcb {\mbox{LHCb}\xspace}

\def\MagUp {\mbox{\em Mag\kern -0.05em Up}\xspace}

\ifthenelse{\boolean{uprightparticles}}%
{

 \def\PDelta      {\ensuremath{\Delta}\xspace}                 
 \def\PXi      {\ensuremath{\Xi}\xspace}                 
 \def\PLambda      {\ensuremath{\Lambda}\xspace}                 
 \def\PSigma      {\ensuremath{\Sigma}\xspace}                 
 \def\POmega      {\ensuremath{\Omega}\xspace}                 
 \def\PUpsilon      {\ensuremath{\Upsilon}\xspace}                 
 
 \def\PB      {\ensuremath{\mathrm{B}}\xspace}                 
                  
 \def\PD      {\ensuremath{\mathrm{D}}\xspace}

 \def\PK      {\ensuremath{\mathrm{K}}\xspace}

 \def\Pi      {\ensuremath{\mathrm{i}}\xspace}

}
{

 \mathchardef\PDelta="7101
 \mathchardef\PXi="7104
 \mathchardef\PLambda="7103
 \mathchardef\PSigma="7106
 \mathchardef\POmega="710A
 \mathchardef\PUpsilon="7107
                  
 \def\PB      {\ensuremath{B}\xspace}                 
                  
 \def\PD      {\ensuremath{D}\xspace}

 \def\PK      {\ensuremath{K}\xspace}

 \def\Pi      {\ensuremath{i}\xspace}

}

\DeclareRobustCommand{\optbar}[1]{\shortstack{{\miniscule (\rule[.5ex]{1.25em}{.18mm})}
  \\ [-.7ex] $#1$}}

  \def\Kbar    {{\kern 0.2em\overline{\kern -0.2em \PK}{}}\xspace}

\def\KorKbar    {\kern 0.18em\optbar{\kern -0.18em K}{}\xspace}

  \def\Dbar    {{\kern 0.2em\overline{\kern -0.2em \PD}{}}\xspace}

\def\DorDbar    {\kern 0.18em\optbar{\kern -0.18em D}{}\xspace}

\def\Bbar    {{\ensuremath{\kern 0.18em\overline{\kern -0.18em \PB}{}}}\xspace}

\def\BorBbar    {\kern 0.18em\optbar{\kern -0.18em B}{}\xspace}

  \def\Y#1S{\ensuremath{\PUpsilon{(#1S)}}\xspace}

\def\Lbar        {{\ensuremath{\kern 0.1em\overline{\kern -0.1em\PLambda}}}\xspace}
\def\LorLbar    {\kern 0.18em\optbar{\kern -0.18em \PLambda}{}\xspace}

\def\AT#1     {\ensuremath{A_{\mathrm{T}}^{#1}}\xspace}           

\def\C#1      {\ensuremath{\mathcal{C}_{#1}}\xspace}                       
\def\Cp#1     {\ensuremath{\mathcal{C}_{#1}^{'}}\xspace}                    
\def\Ceff#1   {\ensuremath{\mathcal{C}_{#1}^{\mathrm{(eff)}}}\xspace}        
\def\Cpeff#1  {\ensuremath{\mathcal{C}_{#1}^{'\mathrm{(eff)}}}\xspace}       
\def\Ope#1    {\ensuremath{\mathcal{O}_{#1}}\xspace}                       
\def\Opep#1   {\ensuremath{\mathcal{O}_{#1}^{'}}\xspace}                    

\newcommand{\tev}{\ifthenelse{\boolean{inbibliography}}{\ensuremath{~T\kern -0.05em eV}\xspace}{\ensuremath{\mathrm{\,Te\kern -0.1em V}}}\xspace}
\newcommand{\gev}{\ensuremath{\mathrm{\,Ge\kern -0.1em V}}\xspace}
\newcommand{\mev}{\ensuremath{\mathrm{\,Me\kern -0.1em V}}\xspace}
\newcommand{\kev}{\ensuremath{\mathrm{\,ke\kern -0.1em V}}\xspace}
\newcommand{\ev}{\ensuremath{\mathrm{\,e\kern -0.1em V}}\xspace}
\newcommand{\gevc}{\ensuremath{{\mathrm{\,Ge\kern -0.1em V\!/}c}}\xspace}
\newcommand{\mevc}{\ensuremath{{\mathrm{\,Me\kern -0.1em V\!/}c}}\xspace}
\newcommand{\gevcc}{\ensuremath{{\mathrm{\,Ge\kern -0.1em V\!/}c^2}}\xspace}
\newcommand{\gevgevcccc}{\ensuremath{{\mathrm{\,Ge\kern -0.1em V^2\!/}c^4}}\xspace}
\newcommand{\mevcc}{\ensuremath{{\mathrm{\,Me\kern -0.1em V\!/}c^2}}\xspace}

\def\mm   {\ensuremath{\rm \,mm}\xspace}

\def\mum  {\ensuremath{{\,\upmu\rm m}}\xspace}

\def\invfb   {\ensuremath{\mbox{\,fb}^{-1}}\xspace}

\def\gsim{{~\raise.15em\hbox{$>$}\kern-.85em
          \lower.35em\hbox{$\sim$}~}\xspace}
\def\lsim{{~\raise.15em\hbox{$<$}\kern-.85em
          \lower.35em\hbox{$\sim$}~}\xspace}

\def\ponn {\ensuremath{\rm {\it{p^+}}\mbox{-}on\mbox{-}{\it{n}}}\xspace}

\def\tell1  {TELL1\xspace}
\def\ukl1   {UKL1\xspace}

\newcommand{\eg}{\mbox{\itshape e.g.}\xspace}
\newcommand{\ie}{\mbox{\itshape i.e.}\xspace}

\newcommand{\cf}{\mbox{\itshape cf.}\xspace}

\usepackage{cite} 
\usepackage{mciteplus}

\usepackage{longtable} 

\usepackage{tabularx} 
\usepackage{multicol} 
\usepackage{booktabs} 
\usepackage{caption}
\usepackage{bm} 

\usepackage{overpic} 

\def\Put(#1,#2)#3{\leavevmode\makebox(0,0){\put(#1,#2){#3}}}

\begin{document}
\renewcommand{\thefootnote}{\fnsymbol{footnote}}
\setcounter{footnote}{1}

\begin{titlepage}
\pagenumbering{roman}


\vspace*{-1.5cm}
\centerline{\large EUROPEAN ORGANIZATION FOR NUCLEAR RESEARCH (CERN)}
\vspace*{1.15cm}
\hspace*{-0.5cm}
\begin{tabular*}{\linewidth}{lc@{\extracolsep{\fill}}r}
  \vspace*{-2.7cm}\mbox{\!\!\!\includegraphics[width=.14\textwidth]{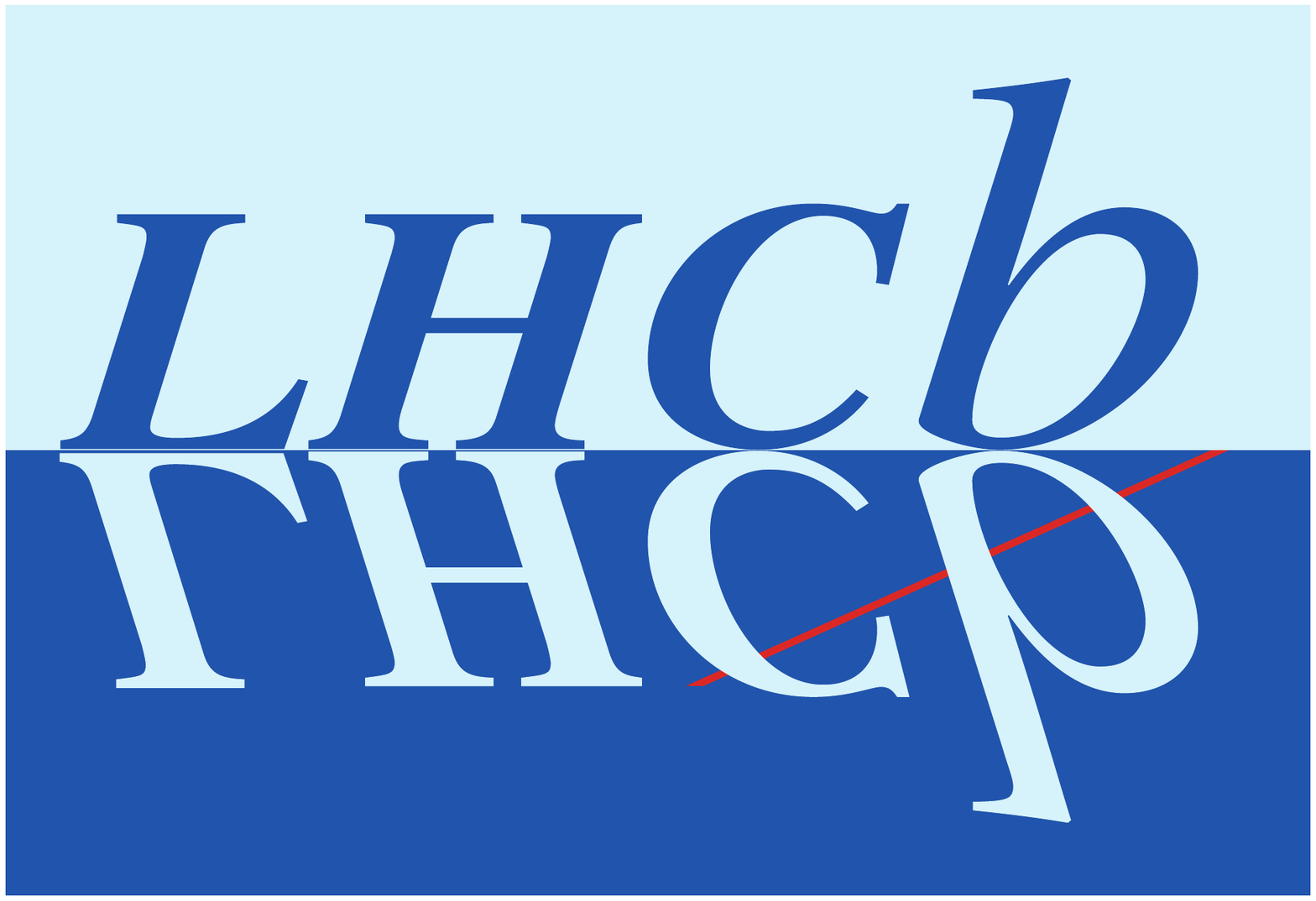}}
  & &%
  \\ & & CERN-LHCb-DP-2018-003
  \\ & & August 26, 2020
  \\ & & \\
\end{tabular*}

\vspace*{0.7cm}


{\bf\boldmath\huge
  \begin{center}
    Monitoring radiation damage in the LHCb Tracker Turicensis
  \end{center}
  }

\vspace*{0.5cm}


\begin{center}
  The LHCb Silicon Tracker group
 \bigskip

C.~Abellan~Beteta$^{1}$,
M.~Atzeni$^{1}$,
V.~Battista$^{2}$,
A.~Bursche$^{9}$,
B.~Dey$^{10}$,
A.~Dosil~Suarez$^{4}$,
C.~Elsasser$^{1}$,
A.~Fernandez~Prieto$^{4}$,
J.~Fu$^{3}$,
E.~Graverini$^{2}$,
I.~Komarov$^{2,b}$,
E.~Lemos~Cid$^{4}$,
F.~Lionetto$^{1}$,
A.~Mauri$^{7}$,
A.~Merli$^{3}$,
P.R.~Pais$^{6}$,
E.~Perez~Trigo$^{4}$,
M.~del~Pilar~Peco~Regales$^{5,a}$,
P.~Stefko$^{2}$,
O.~Steinkamp$^{1}$,
B.~Storaci$^{1}$,
M.~Tobin$^{8}$,
A.~Weiden$^{1}$
.\bigskip\newline{\it
\footnotesize
$ ^{1}$Physik-Institut, Universit{\"a}t Z{\"u}rich, Z{\"u}rich, Switzerland\\
$ ^{2}$Institute of Physics, Ecole Polytechnique  F{\'e}d{\'e}rale de Lausanne (EPFL), Lausanne, Switzerland\\
$ ^{3}$Universit{\`a} degli Studi di Milano, Milano, Italy\\
$ ^{4}$Instituto Galego de F{\'\i}sica de Altas Enerx{\'\i}as (IGFAE), Universidade de Santiago de Compostela, Santiago de Compostela, Spain\\
$ ^{5}$Max-Planck-Institut f{\"u}r Kernphysik (MPIK), Heidelberg, Germany\\
$ ^{6}$European Organization for Nuclear Research (CERN), Geneva, Switzerland\\
$ ^{7}$Nikhef National Institute for Subatomic Physics, Amsterdam, Netherlands\\
$ ^{8}$Chinese Academy of Sciences, Beijing, China\\
$ ^{9}$Sezione INFN di Cagliari, Cagliari, Italy\\
$ ^{10}$Institute of Particle Physics, Central China Normal University, Wuhan, Hubei, China\\
\bigskip
$ ^{a}$now at AGH University of Science and Technology, Krakow, Poland\\
$ ^{b}$now at Deutsches Elektronen-Synchrotron, Hamburg, Germany
}

\end{center}

\vspace*{0.5cm}


\begin{abstract}
\noindent This paper presents the techniques used to monitor radiation damage in the LHCb
Tracker Turicensis during the LHC Runs 1 and 2. Bulk leakage currents in
the silicon sensors were monitored continuously, while the full depletion
voltages of the sensors were estimated at regular intervals by performing
dedicated scans of the charge collection efficiency as a function of the
applied bias voltage. Predictions of the expected leakage currents and full
depletion voltages are extracted from
the simulated radiation profile, the luminosity delivered by the LHC, and the
thermal history of the silicon sensors. Good agreement between measurements
and predictions is found.
  
\end{abstract}

 \vspace*{0.7cm}

 \begin{center}
   Published in JINST 15, P08016
 \end{center}

 \vspace{\fill}

 {\footnotesize 
   \centerline{ 
     \copyright~CERN on behalf of the \lhcb collaboration, license
     \href{http://creativecommons.org/licenses/by/4.0/}{CC-BY-4.0}.
     }
   }

\vspace*{-10mm}

\end{titlepage}

\cleardoublepage

\setcounter{page}{2}

\renewcommand{\thefootnote}{\arabic{footnote}}
\setcounter{footnote}{0}

\pagestyle{plain}
\setcounter{page}{1}
\pagenumbering{arabic}

\newcommand{\feq}{\ensuremath{\Phi_\text{1 MeV-n,eq}}\xspace}
\newcommand{\neff}{\ensuremath{n_\text{eff}}\xspace}

\section{Introduction \label{sec:Intro}}

The LHCb experiment~\cite{Alves:2008zz} at the LHC was designed to study decays of hadrons containing beauty and charm quarks. 
Its main goal is to perform precision measurements of $C\!P$ violating observables, and to search for signatures of potential physics beyond the Standard Model in rare decays of such hadrons.

\begin{figure}[tb]
  \centering
  \includegraphics[width=0.7\textwidth]{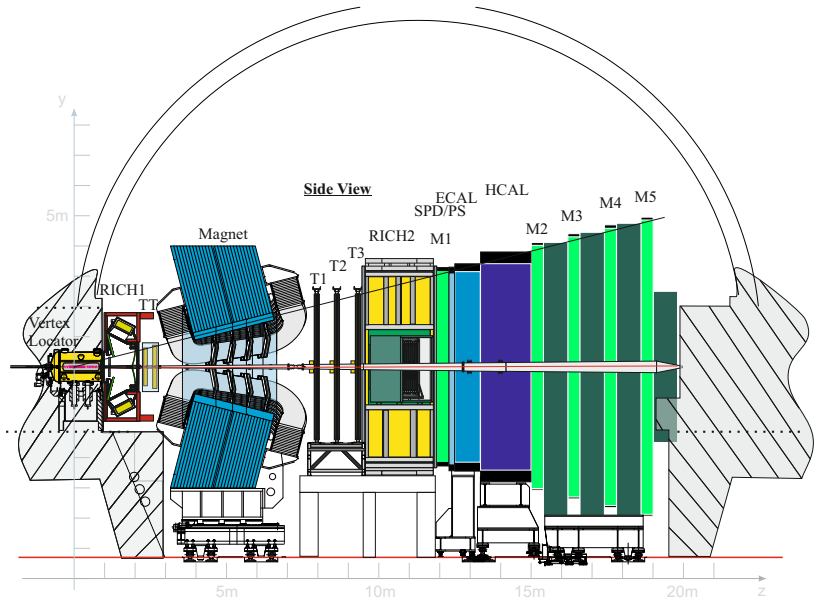}
  \caption{Layout of the LHCb detector: vertex locator (VELO), Ring-Image Cherenkov counters (RICH1 and RICH2), tracking stations upstream (TT) and downstream (T1-T3) of the spectrometer dipole magnet, scintillating pad detector used for trigger purposes (SPD), pre-shower detector (PS) and electromagnetic calorimeter (ECAL), hadronic calorimeter (HCAL), and muon stations (M1-M5). LHCb uses a right-handed cartesian coordinate system with its origin at the nominal interaction point, the $z$-axis pointing along the LHC beam axis and the $y$ axis pointing upwards.}
    \label{fig:lhcb}
\end{figure}

The layout of the LHCb detector during the LHC Runs~1 and~2 is illustrated in Fig.~\ref{fig:lhcb}. 
The detector is a single-arm forward spectrometer that covers polar angles from about 15~mrad to 250~mrad and consists of a vertex locator, a tracking system comprising a single planar tracking station upstream of the spectrometer dipole magnet and three tracking stations downstream of this magnet, two Ring Imaging Cherenkov detectors, a calorimeter system comprising electromagnetic and hadronic calorimeters, and a muon system.  
In the LHC Run~1, the Silicon Tracker (ST) comprised the tracking station upstream of the dipole magnet, called Tracker Turicensis (TT), and the central part in each of the three tracking stations downstream of the dipole magnet, called Inner Tracker (IT); the outer part of the downstream tracking stations was formed by a straw-tube detector. The tracking system has been dismantled in the context of a substantial detector upgrade for the LHC Run 3. The TT will be replaced by a new silicon strip detector, while the downstream tracker will consist of scintillating fibre planes~\cite{LHCb-TDR-012,*LHCb-TDR-015}.

The TT employed conventional AC-coupled float-zone \ponn silicon micro-strip sensors that are 500~\mum thick and have 512~strips with a pitch of 183~\mum and a length of 96~\mm. 
The total silicon volume per sensor is 4512 mm$^3$.
The initial full depletion voltages of the sensors were determined from measurements of the bulk capacitance as a function of applied bias voltage and were found to range between 135 and 275\,V.

\begin{figure}[tb]
  \centering
  \includegraphics[width=0.7\textwidth]{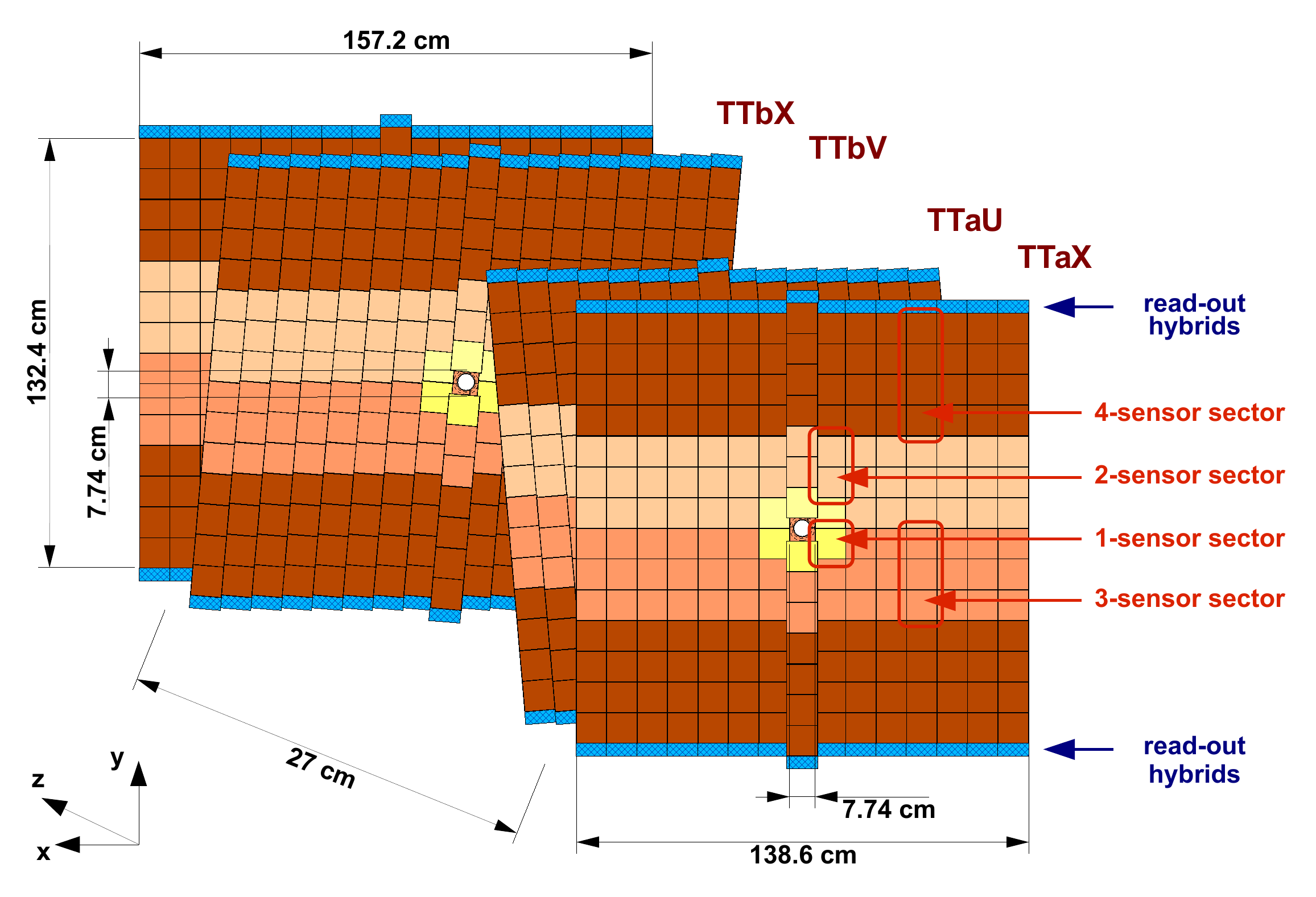}
  \caption{Layout of the TT, with the LHC beam pipe passing through an opening in the centre of the detection layers. The four detection layers are labelled TTaX, TTaU, TTbV and TTbX. The four different types of read-out sectors employed in each of the detection layers are indicated by different shading: read-out sectors close to the beam pipe consist of a single silicon sensor, other read-out sectors consist of two, three or four silicon sensors that are connected together in series.
    \label{fig:TT_layout}
    }
\end{figure}

The TT consisted of one detector box containing four detection layers arranged in two pairs: the first two layers were centered around $z = 232$~cm, and the last two around $z = 262$~cm. 
An ambient temperature of about 8$^\circ$C was maintained inside the box. 
The box was flushed with dry gas ($N_2$) to avoid condensation on cold surfaces. 
Silicon sensors within each detection layer were electronically grouped into read-out sectors consisting of one, two, three or four sensors, as illustrated in Fig.~\ref{fig:TT_layout}. 
All sensors within a read-out sector were connected in series to a front-end read-out hybrid that carried four 128-channel Beetle chips~\cite{Agari:2004ja}.

The Beetle front-end read-out chip amplifies and shapes the signals from the read-out strips and samples the signal amplitude every 25~ns, corresponding to the LHC bunch-crossing frequency of 40~MHz. 
The time offset between the LHC bunch clock and the Beetle sampling time is an adjustable parameter. 
The output data of the Beetle chips are multiplexed, digitised and transmitted optically to the LHCb TELL1 read-out boards~\cite{Haefeli:2006cv}. 
In the TELL1 boards, algorithms for common-mode subtraction, cluster finding and zero-suppression are executed during normal data taking. 
It is, however, also possible to store the raw sampled amplitudes from each input channel of the Beetle chip for offline analysis. 
This non-zero-suppressed read-out mode was employed for the studies discussed in this paper.

\begin{figure}
\centering
\raisebox{-.5\height}{\includegraphics[width=.46\textwidth]{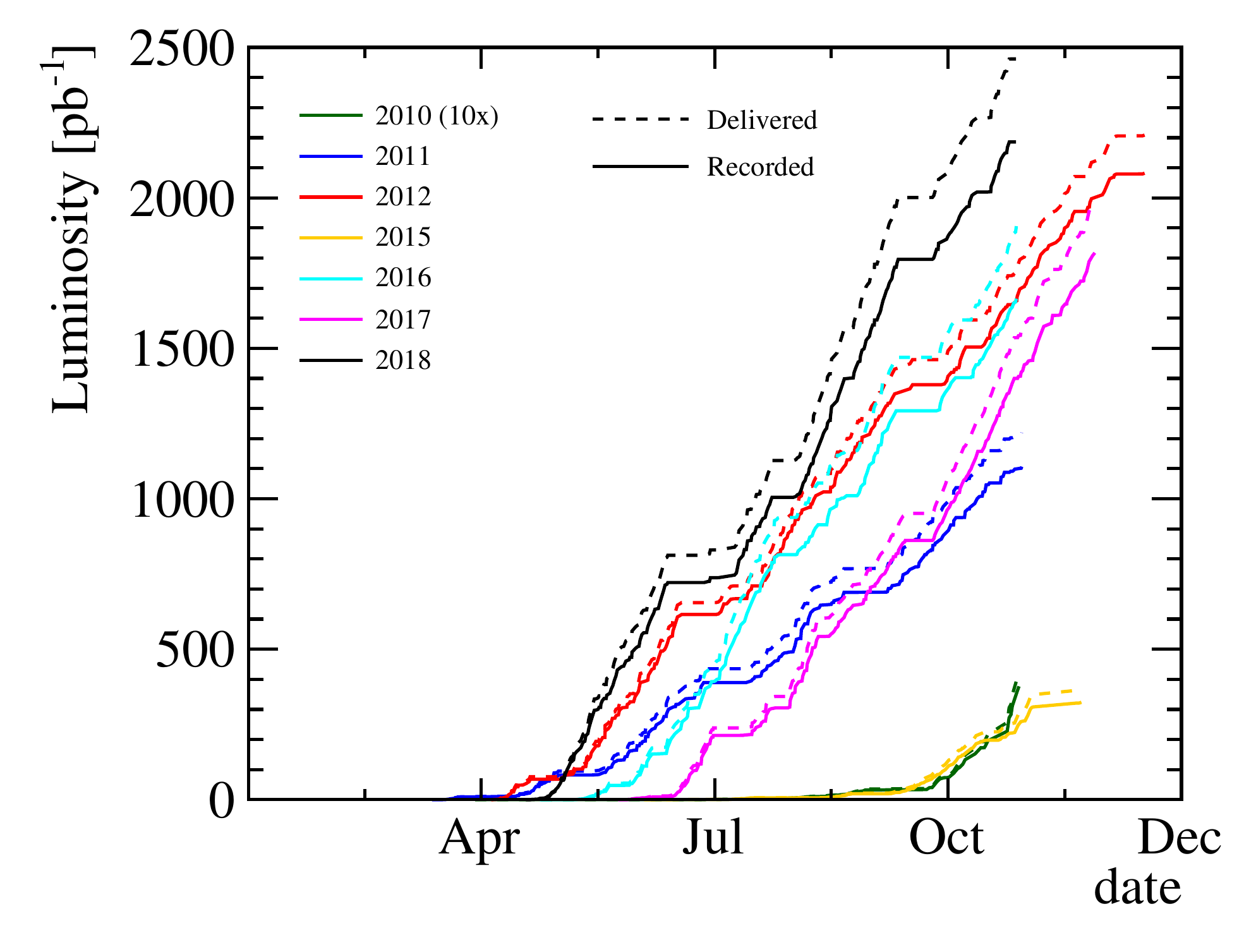}}
\raisebox{-.5\height}{\includegraphics[width=.53\textwidth]{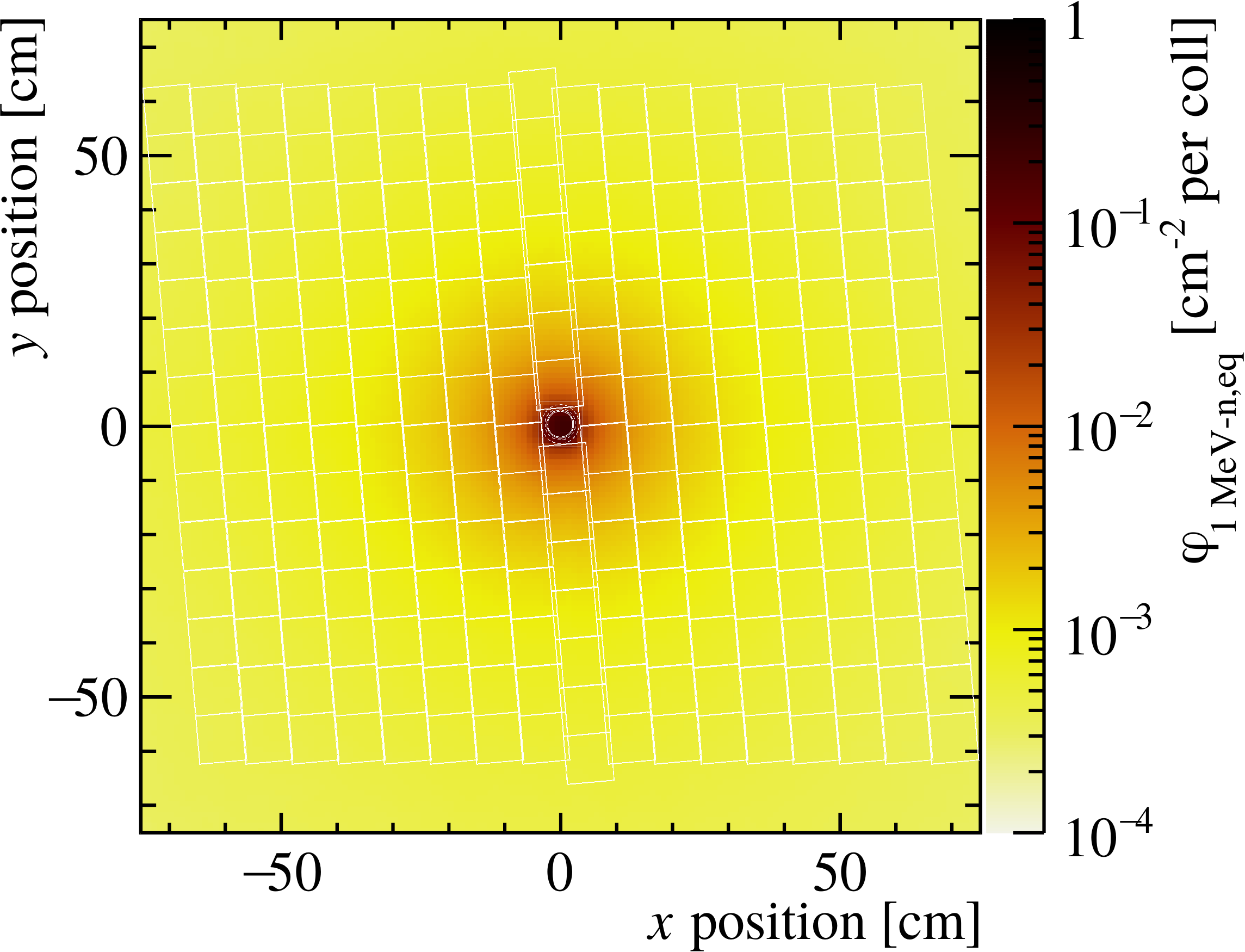}}
\caption{Left: evolution of the integrated luminosity for $pp$ collisions in each year of LHC operation. 
Right: expected 1-\mev neutron equivalent fluence per proton-proton
        collision, with the TTaU detector layer overlaid. These
        results were obtained from a \textsc{Fluka} simulation of
        the LHCb detector, assuming a proton-proton collision energy of 14~TeV.}\label{fig:neutronmap}
\end{figure}

LHCb collected an integrated luminosity of about 3~\invfb at proton-proton collision energies of $\sqrt{s}=$7~and 8~\tev during Run~1 of the LHC (2010--2012), and about 6~\invfb at 13~\tev in Run~2 (2015--2018).
The evolution of the integrated luminosity is shown in Fig.~\ref{fig:neutronmap} (left) for each year of data taking.
Expected particle fluences at collision center-of-mass energies of  $\sqrt{s}=7$, $8$ and $14$~\tev have been derived from simulation using \textsc{Fluka}~\cite{Ferrari:2005zk,*Battistoni:2007zzb}, a general purpose tool for calculations of particle transport and interactions with matter, and a detailed description of the geometry of the LHCb experimental area~\cite{Karacson:2243499, Corti:1038944}.
Maps of the particle fluence at $\sqrt{s}=7$ and $8$~TeV are scored with a resolution of $[1 ,1 ]$~cm, while the map at $\sqrt{s}=14 $~\tev is scored with a resolution of $[2.5 ,2.5 ]$~cm and subsequently interpolated to obtain a map with a resolution of $[1 ,1 ]$~cm. All maps are produced at a $z$-coordinate corresponding to the location of the TTaU layer.
The statistical uncertainty on the expected fluence is modeled as a power-law function of the radial position from the beam pipe, $a_{0} \times (r/r_{0})^\alpha$, where $a_{0} = 0.7\times10^{-2}$ cm$^{-2}$ , $r_{0} = 1$ cm, and $\alpha$ = 0.918.
The expected 1-\mev neutron equivalent fluence per proton-proton collision at a center-of-mass energy of $\sqrt{s}=14$~\tev at the location of one of the detection layers of the TT is shown in Figure~\ref{fig:neutronmap} (right).
The highest expected 1-\mev neutron equivalent fluence, in the innermost region of the TT, corresponds to about $10^{13}$\,cm$^{-2}$ per $1\,\invfb$ of integrated luminosity collected at a proton-proton collision energy of $\sqrt{s}=14$~\tev. 
Since particles are primarily produced in the forward direction, the expected fluence falls off by almost three orders of magnitude from the innermost to the outermost region of the TT detector. 

Two methods were employed to monitor the radiation damage to the silicon sensors and its impact on the detector performance. 
The first method, discussed in Section~\ref{sec:ileak}, is based on the continuous measurement of detector leakage currents. 
These measurements are compared to the expected radiation-induced increase of the bulk leakage current as a function of received fluence at an ambient temperature of $8^{\circ}$~C.
The second method, discussed in Section~\ref{sec:vdepl}, is based on the analysis of collision data accumulated during dedicated Charge Collection Efficiency (CCE) scans. 
These CCE scans were performed at regular intervals and allow an estimation of the full depletion voltage of the sensors. 
The results of these measurements are compared to predictions derived from the Hamburg model~\cite{Moll:1999kv}, which parametrises the evolution of the effective doping concentration of the silicon bulk as a function of received fluence and temperature.
Section~\ref{sec:summary} presents a summary of the different studies and discusses the expected evolution of the TT performance in terms of radiation damage.

\section{Leakage current measurements\label{sec:ileak}}

Bias voltage for the TT was supplied by a commercial high-voltage (HV) system\footnote{CAEN SY1527 crates with A1511B modules, by CAEN S.p.A., Viareggio, Italy.} with 152~channels supplying up to 500~V at a maximum current of 10~mA per HV channel. 
The innermost six silicon sensors in each detection layer, which received the highest fluence and therefore experienced the largest increase in leakage current, were connected to individual HV channels. 
In the outer regions of the detection layers, groups of two, three, four, nine or twelve silicon sensors were connected to a common HV channel. 
The current drawn by each HV channel was monitored with a resolution of 1.0 $\mu$A, and a maximum interval of 120 minutes allowed between consecutive readings. 
For the purpose of this analysis, the leakage current, $I_\text{leak}$, for a given HV channel is determined for each LHC fill as the maximum current observed in that fill\footnote{A typical LHC fill is several hours in duration.}.
Measurements from fills with special detector configurations (\eg higher or lower bias voltage) are discarded in the leakage current measurements. 
The leakage current has a temperature dependent behaviour given by
\begin{equation}
\frac{I_\text{leak}(T_1)}{I_\text{leak}(T_2)} = \Bigg(\frac{T_1}{T_2}\Bigg)^2\exp\Bigg[\frac{E_{eff}}{2k_B}\Bigg(\frac{1}{T_2}-\frac{1}{T_1}\Bigg)\Bigg],
\end{equation}
where $T_{1,2}$ are different temperatures, $k_B$ is the Boltzmann constant and $E_{eff}=1.21$ eV is the effective energy for the temperature dependence of the bulk current in irradiated silicon sensors, shown to be temperature-independent in Ref.~\cite{Chilingarov_2013}.
All measured currents are normalised to a nominal temperature of 8$^{\circ}$C, using measurements of the ambient temperature inside the detector box and the well-known temperature dependence of the bulk leakage current in silicon~\cite{Sze:2006}.

\begin{figure}[tb]
\begin{center}
{\includegraphics[width=0.85\linewidth]{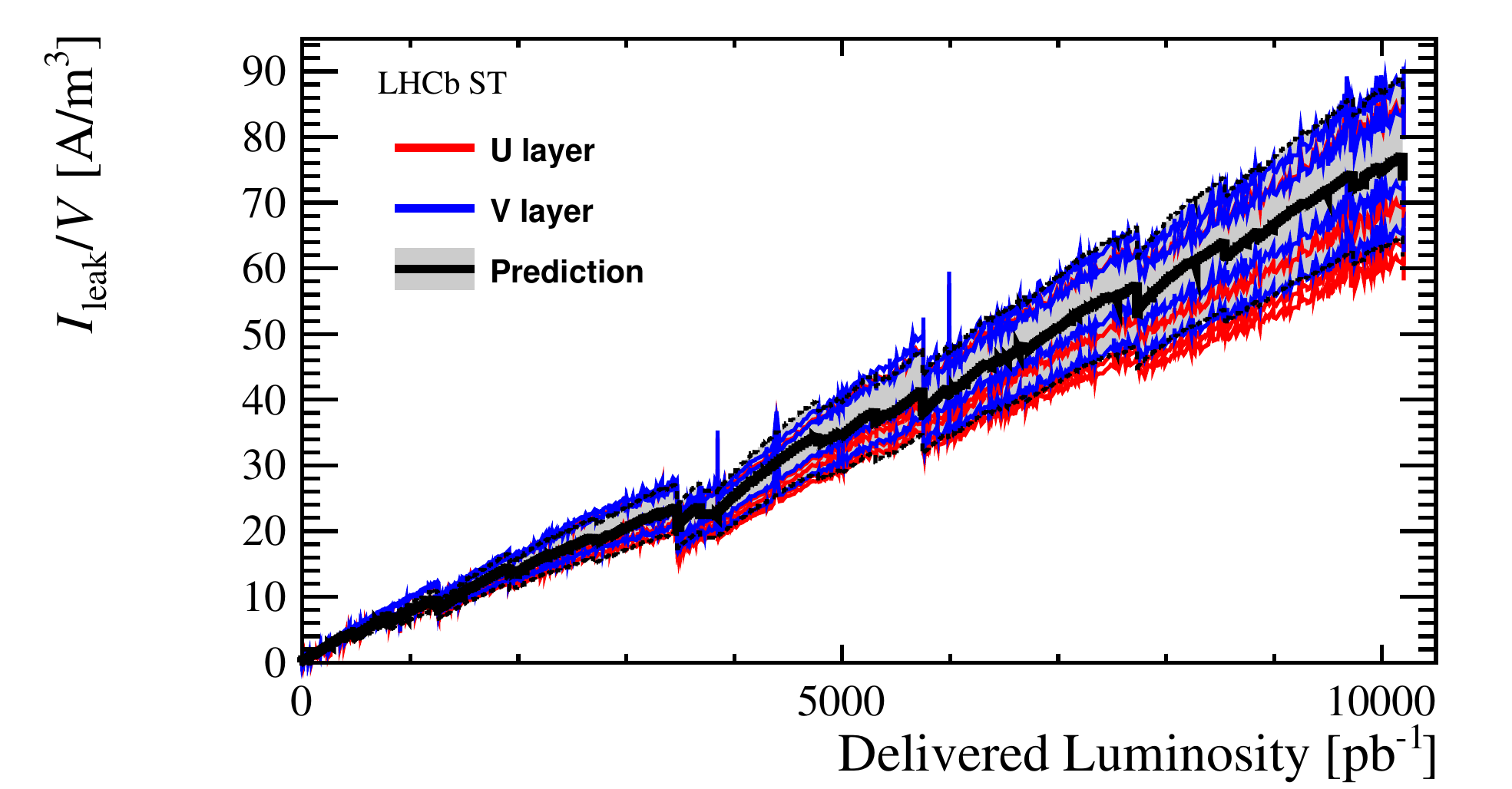}}
{\includegraphics[width=0.85\linewidth]{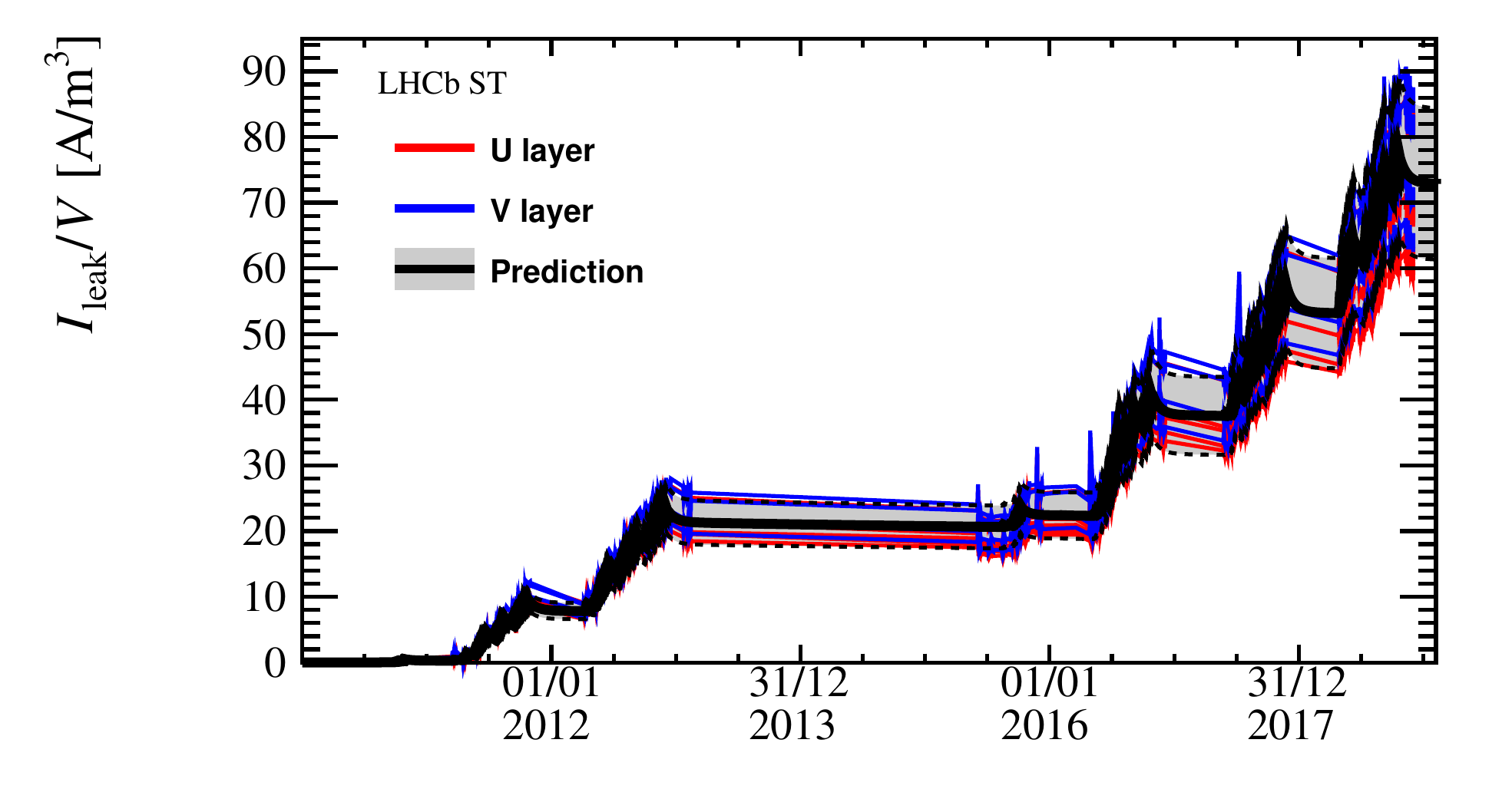}}
\caption[TT leakage current]{Evolution of the leakage current for the different HV channels connected to one-sensor read-out sectors in TT as a function of (a) the delivered integrated luminosity measured and (b) time. The red curves show channels in the detector layer TTaU, the blue one in TTbV. The predicted evolution is shown in black while the grey band shows its uncertainty, computed from the uncertainty on the model parameters, on the \textsc{Fluka} simulation and on the temperature measurements. The uncertainty does not account for the range of fluence expected across the sectors shown.} \label{fig:leakTT}
\end{center}
\end{figure}

The normalized leakage currents, scaled to 8$^{\circ}$C and per silicon volume, are shown as a function of integrated delivered luminosity and time in Fig.~\ref{fig:leakTT} for the innermost sensors in the second and third detection layers, indicated as U and V layers, respectively.

\begin{table}[tb]
\begin{center}
\caption[Parameters for the leakage current predictions]{Parameters for the leakage current predictions.}
\begin{tabularx}{\textwidth}{p{4.5cm}r@{}l@{}l}
\toprule
\textbf{Parameter} & \multicolumn{3}{c}{\textbf{Value}}\\
\midrule
$\alpha_{I,0}$ & $(6.67\pm0.09)$ & $\times 10^{-17}$ & A/cm\\
$\alpha_{I,1}$ & $(7.23\pm0.06)$ & $\times 10^{-17}$ & A/cm\\
$\alpha_{I,2}$ & $(3.08\pm0.07)$ & $\times 10^{-18}$ & A/cm\\
$k_0$          & $(4.2\enspace\pm0.5\enspace)$ & $\times 10^{13}$ & s$^{-1}$\\
$E_a$          & $(1.11\pm0.05)$ &  & eV\\
\bottomrule
\end{tabularx}
\label{tab:leakCurParams}
\end{center}
\end{table}

The change $\Delta I_\text{leak}$ in the leakage current is expected to be linear as a function of the fluence, $\Phi$, that the sensor has been exposed to, for a range between $10^{11}$ and $10^{15}$ 1-MeV n / cm$^2$~\cite{Moll:1999kv}. 
So $\Delta I_\text{leak}$ can be described as
\begin{equation}
\Delta I_\text{leak} = \alpha\cdot\Phi\cdot V,
\end{equation}
where $\alpha$ is the current-related damage rate and $V$ the silicon volume. 
The coefficient $\alpha$ shows a behaviour dependent on temperature, $T_a$, and on time, $t$, as the irradiation occurs alternating with annealing, \ie periods without collisions~\cite{Moll:2002tn}. Short term annealing (on the order of a few days to a month) is described by the sum of a constant and exponential terms, where the first term describes the effects of stable damage. An additional logarithmic term is used to describe annealing effects for time periods longer than a year~\cite{Moll:1999nh},
\begin{equation}
\alpha(t,T_a) = \alpha_{I,0}+\alpha_{I,1}\exp\Big(-t/\tau_I(T_a)\Big)+\alpha_{2}\log\Big(t/t_{0}\Big).
\label{eq:Alpha}
\end{equation}
The temperature dependence of the time constant $\tau_I$ in Eq.~(\ref{eq:Alpha}) can be described by the Arrhenius relation~\cite{Arrhenius:1889}, 
\begin{equation}
\tau_I(T_a) = \Bigg[k_{I,0}\exp\Bigg(-\frac{E_{a,I}}{k_BT_a}\Bigg)\Bigg]^{-1},
\label{eq:Arrhenius}
\end{equation}
where $k_{I,0}$ is the frequency factor and $E_{a,I}$ the activation energy of the process.
The values of these parameters are listed in Table~\ref{tab:leakCurParams}.

Figure~\ref{fig:leakTT} also shows the predicted evolution of $I_\text{leak}$ based on this formalism using the actual running conditions in LHCb (\ie instantaneous luminosity, duration of the fills, ambient temperature) and the 1-\mev neutron equivalent fluence computed for a given sector by integrating the simulated \textsc{Fluka} radiation map over the sensor area.  

The predictions are in good agreement with the evolution of the leakage currents in data. 
Significant decreases in the currents are observed, as expected, in the data during periods with no irradiation, \ie in the long shutdown, LS1, of the LHC in 2013 and 2014, extended stops of the accelerator during the winter season and short technical stops throughout the year (one week about every three months).

\section{Depletion voltage measurements\label{sec:vdepl}}
Non-ionizing energy loss (NIEL) damage in the silicon bulk leads to a change in the effective doping concentration, $n_\text{eff}$, of the bulk and therefore to a change in the full depletion voltage $V_\text{depl}$ since the two quantities are proportional to each other according to the relation 
\begin{equation}
V_\text{depl} = \frac{q}{2\varepsilon\varepsilon_0}n_\text{eff}D^2,
\label{eq:vdepl}
\end{equation}
where $q$ is the electron charge, $\varepsilon\varepsilon_0$ the effective permittivity of silicon and $D$ the full thickness of the sensor.

As mentioned in Section~\ref{sec:Intro}, the initial full depletion voltage of all silicon sensors was determined in $C-V$ scans at the foundry and during the construction of the detector. 
For a subset of the installed detectors, the evolution of $V_\text{depl}$ during operation has then been measured from scans of the charge-collection efficiency as a function of the applied bias voltage (CCE scans). 
Such CCE scans require particles from LHC collisions to be recorded using special data taking settings. 
They are performed about two to four times a year, in particular once at the beginning of LHC operation in spring and once at the end of operation before Christmas.
Table~\ref{tab:CCEscans}~summarises the CCE scans performed between 2011 and 2018. Few CCE scans have been removed from the analysis due to operational inconsistencies, and are not listed in the table.

\begin{table}
\caption{Summary of CCE scans dates and corresponding integrated delivered luminosity.}\label{tab:CCEscans}
\centering
    \begin{tabular}{cccc}
        \toprule
        \textbf{Date} & \multicolumn{3}{c}{\textbf{Integrated Luminosity  [fb$\bm{^{-1}}$]}}\\
        & \textbf{7~TeV} & \textbf{8~TeV} & \textbf{13~TeV}\\
        \midrule
        2011-07-14 & 0.48 & -- & --\\
        2011-09-07 & 0.81 & -- & --\\
        2011-10-25 & 1.24 & -- & --\\
        \midrule
        2012-04-05 & 1.26 & 0.01 & --\\
        2012-07-02 & 1.26 & 0.66 & --\\
        2012-09-28 & 1.26 & 1.46 & --\\
        \midrule
        2015-06-03 & 1.26 & 2.21 & 0.00\\
        2015-10-20 & 1.26 & 2.21 & 0.23\\
        2016-04-23 & 1.26 & 2.21 & 0.37\\
        2016-08-04 & 1.26 & 2.21 & 1.34\\
        2017-06-05 & 1.26 & 2.21 & 2.28\\
        2017-07-12 & 1.26 & 2.21 & 2.51\\
        2017-09-12 & 1.26 & 2.21 & 3.21\\
        2017-10-27 & 1.26 & 2.21 & 3.81\\
        2017-11-21 & 1.26 & 2.21 & 4.16\\
        2018-04-17 & 1.26 & 2.21 & 4.26\\
        2018-06-26 & 1.26 & 2.21 & 5.07\\
        2018-09-23 & 1.26 & 2.21 & 6.26\\
        \bottomrule
    \end{tabular}
\end{table}

While in the nominal data taking configuration the applied bias voltage $V_\text{bias}$ was 300~V for all TT read-out sectors, during a CCE scan the $V_\text{bias}$ of one detection layer (TTaU) is scanned in a 60~V to 400~V range.
The other detection layers are operated at the nominal bias voltage. 
The data from these layers and from the other detectors of the LHCb tracking system are used to reconstruct the trajectories of charged particles produced in the proton-proton collisions.

Each CCE scan consists of 66 different data taking configurations. For each of the 11 bias voltage steps, 6 sampling time steps are performed.
As the drift velocity of charge carriers and therefore the signal collection time depends on the applied bias voltage, in fact, a scan of the signal sampling time of the Beetle front-end ASIC is performed for each bias voltage setting in the range $\delta t = -9.38,\,-4.69,\,0.00,\,4.69,\,9.38,\,14.06$~ns with respect to the nominal sampling time.
A minimum number of 500000 events are recorded for each calibration step, which roughly corresponds to a total of 2--3 hours of data taking depending on the beam configuration at the time of the CCE scan. These scans are mainly taken during the beam commissioning phase, when only a small number of proton bunches circulate in the accelerator, in order to minimize the loss of useful data for physics analysis.

\begin{figure}[tb]
\begin{center}
\begin{minipage}[t]{0.49\textwidth}
\begin{overpic}[width=\textwidth,scale=.25,tics=20]{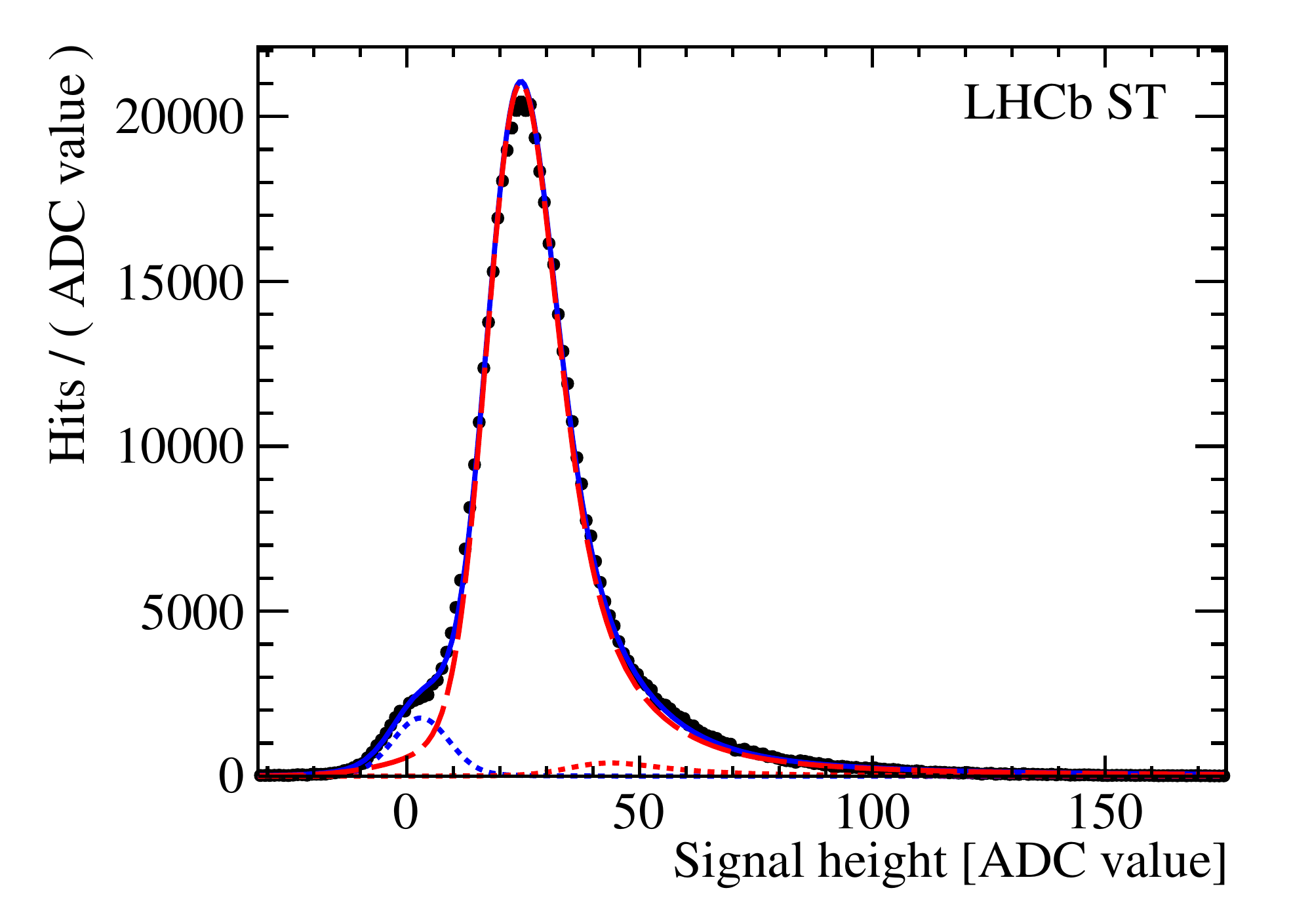}
\put(52,55){(a) $V_\text{bias}=100$~V}
\end{overpic}
\end{minipage}
\begin{minipage}[t]{0.49\textwidth}
\begin{overpic}[width=\textwidth,scale=.25,tics=20]{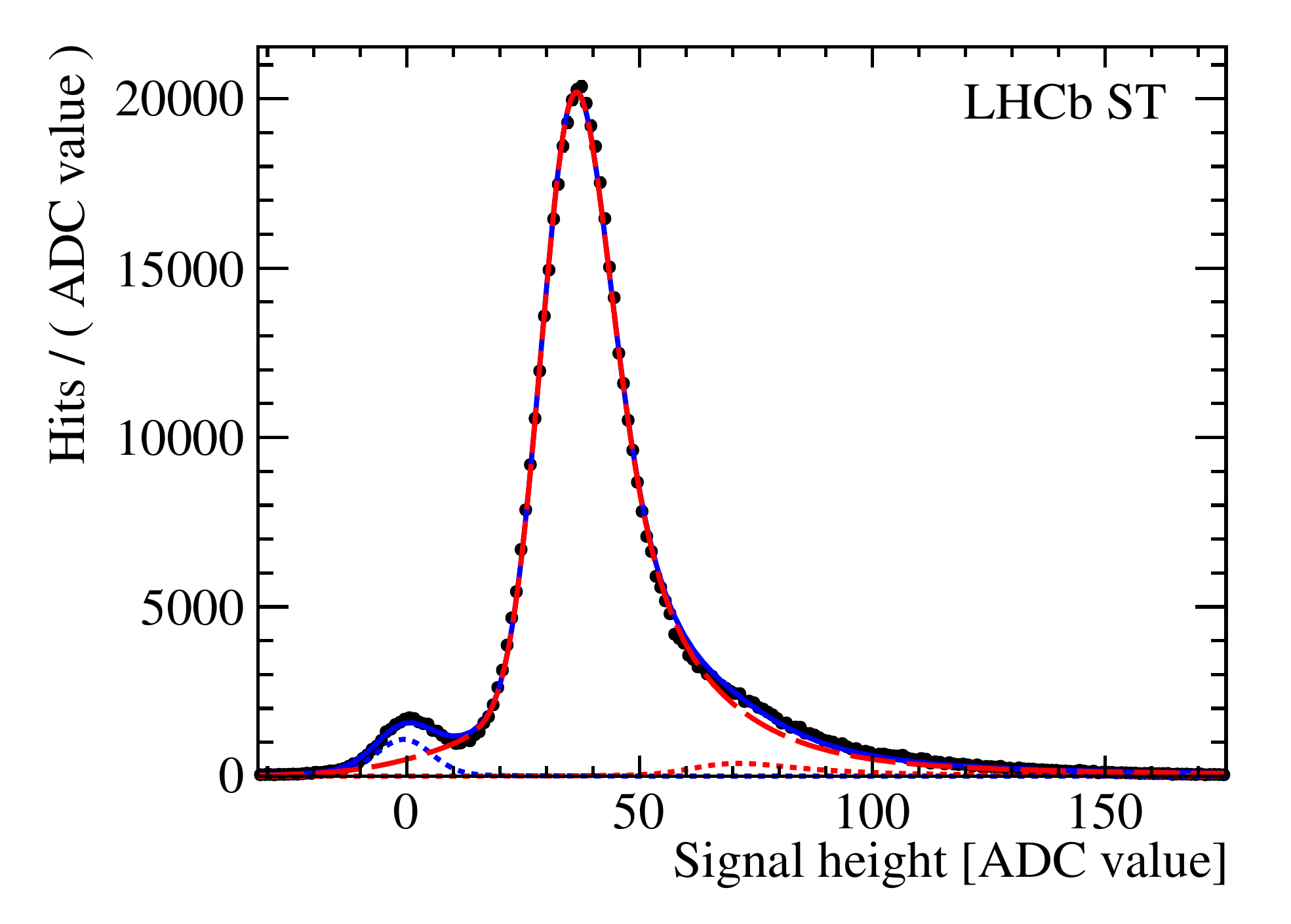}
\put(52,55){(b) $V_\text{bias}=400$~V}
\end{overpic}
\end{minipage}
\caption[Signal height distribution of hits associated to tracks]{Signal height distribution for hits associated to tracks in the TT for two different bias voltage values. The fitted distribution is described in the text.}
\label{fig:signalheightCCE}
\end{center}
\end{figure}

Tracks reconstructed by the rest of the LHCb tracking system, including the TT layers operating in nominal conditions, are used to interpolate the hit position in the scanned layer.
Track quality requirements are applied to remove tracks with poor track fit quality or originating from random combinations of track segments from different parts of the tracking system. 
The pedestal-subtracted ADC values of the three read-out strips closest to this interpolated hit position are then summed to give an estimate of the collected charge.
As an example, Fig.~\ref{fig:signalheightCCE} shows the signal height distribution in a TT read-out sector for $V_\text{bias}=100$~V and $400$~V with $\delta t = 0.00$~ns. 
The incomplete depletion of the silicon bulk at $V_\text{bias}=100$~V induces the clearly visible decrease in signal amplitude.

The most probable value (MPV) of the collected charge is extracted from a fit of this distribution, using a model consisting of the sum of two Gaussians to describe the residual contribution of noise hits due to wrong track extrapolations (blue line), convolved with the sum of two Landau distributions to describe the contribution from signal hits (red line). 
The second Landau distribution describes a signal component due to photon conversions in the beam pipe and detector material upstream of the TT, creating $e^+e^-$ pairs that will cross the TT at the same position and deposit twice as much charge as a single track. 
The MPV and the width of this second Landau distribution are fixed to twice the corresponding values of the first Landau.

\begin{figure}[tb]
\begin{center}
\begin{minipage}[t]{0.49\textwidth}
\begin{overpic}[width=\textwidth,scale=.25,tics=20]{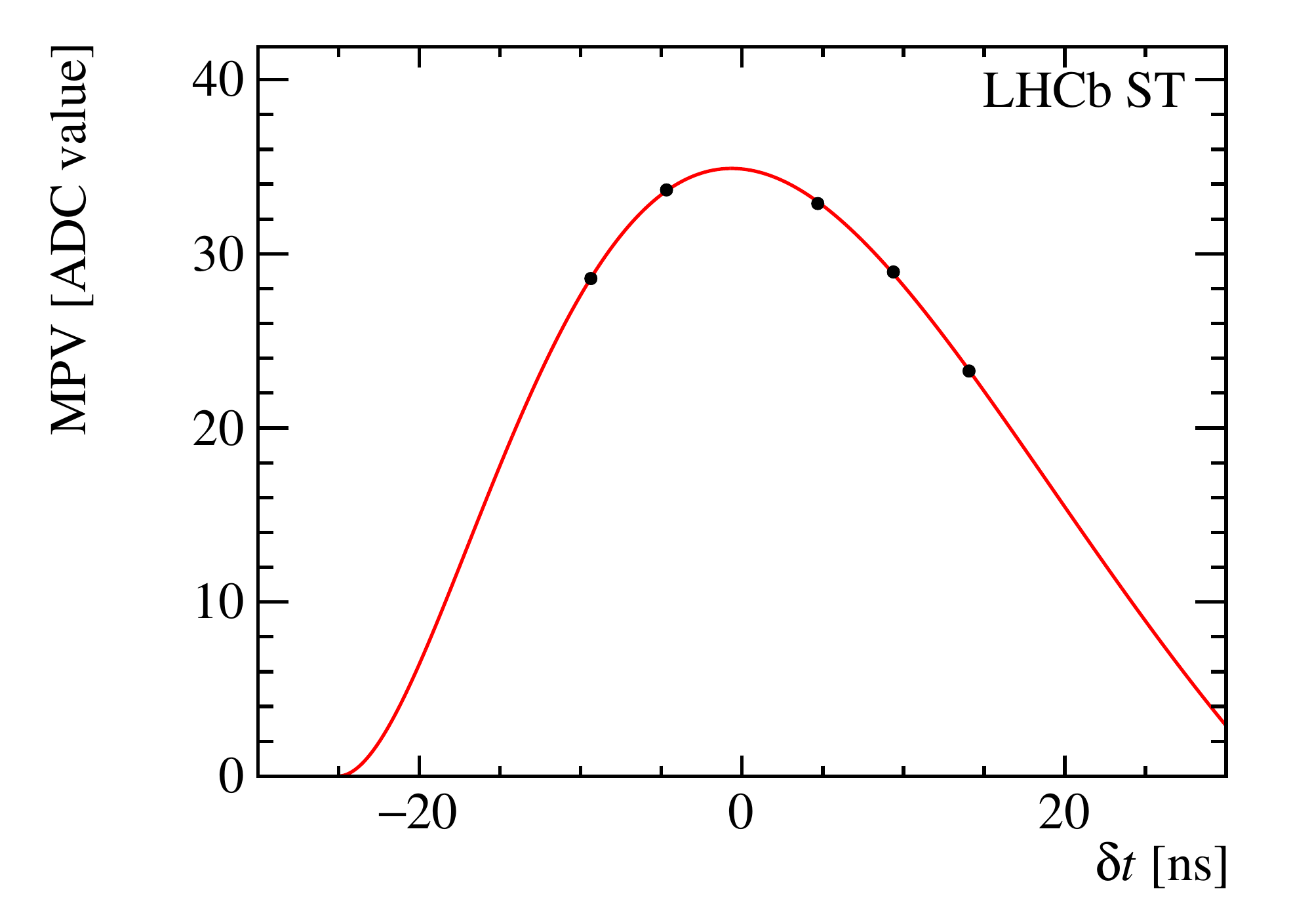}
\put(43,15){(a) $V_\text{bias}=100$~V}
\end{overpic}
\end{minipage}
\begin{minipage}[t]{0.49\textwidth}
\begin{overpic}[width=\textwidth,scale=.25,tics=20]{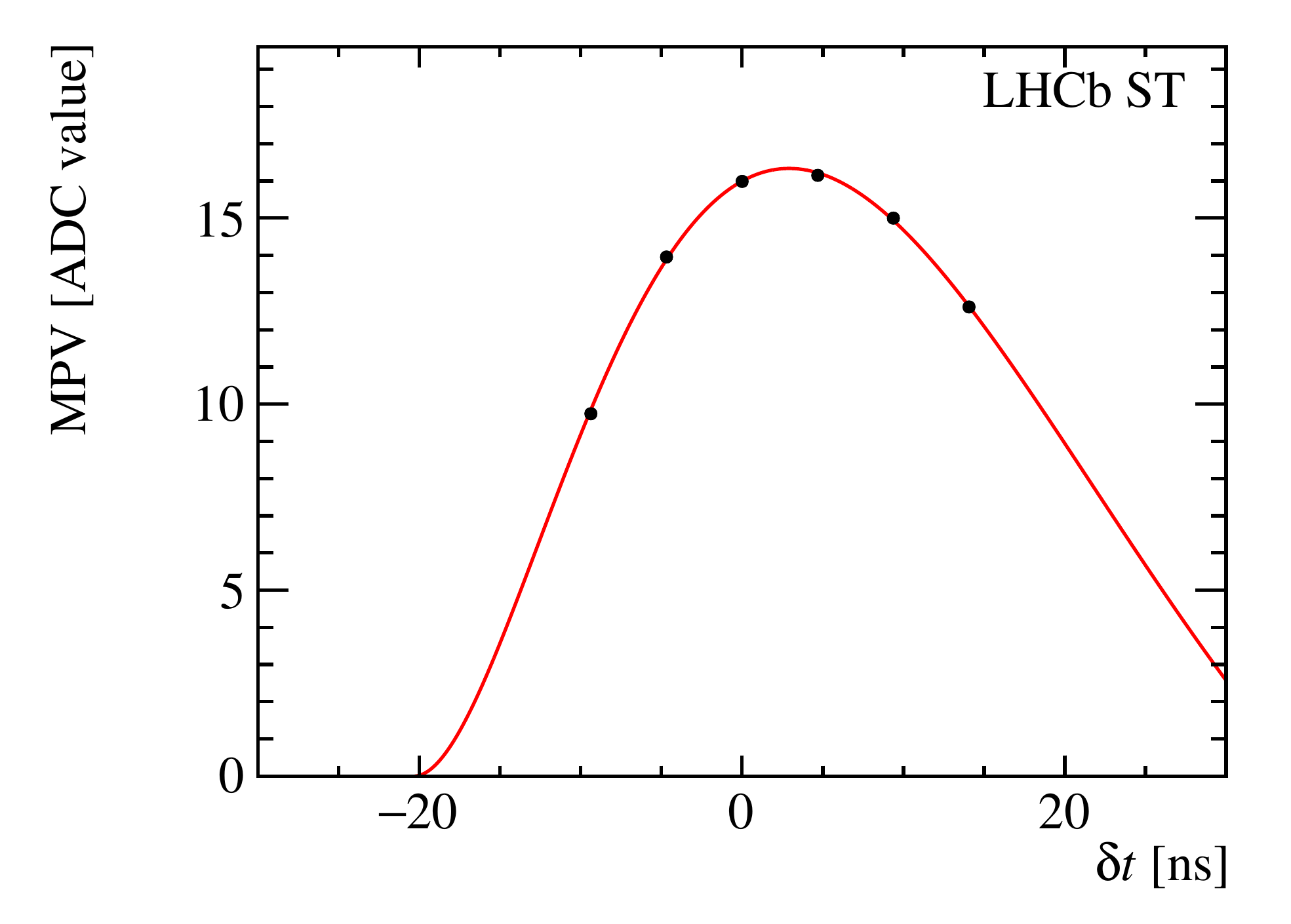}
\put(43,15){(b) $V_\text{bias}=400$~V}
\end{overpic}
\end{minipage}
\caption[Pulse shape fit]{Pulse shapes recorded in the TT for two different bias voltage values. The fitted function is shown in Eq.~(\ref{eq:pulse}).}
\label{fig:pulseCCE}
\end{center}
\end{figure}

The MPVs extracted for different sampling times $\delta t$ at a given bias voltage are then fit with the empirical function 
\begin{equation}
f(\delta t|\tau,t_0,A) = A\cdot\exp\Bigg(-\frac{\delta t-t_0}{\tau}\Bigg)\cdot\Bigg[\frac{1}{2}\Bigg(\frac{\delta t-t_0}{\tau}\Bigg)^2-\frac{1}{6}\Bigg(\frac{\delta t-t_0}{\tau}\Bigg)^3\Bigg]
\label{eq:pulse}
\end{equation}
describing the semi-Gaussian signal shape output by the pulse shaping circuit employed in the Beetle chip, which is composed of a differentiator (CR) followed by an integrator (RC)~\cite{Lochner:2006sq,Elsasser:2002199}. 
The shaping time $\tau$, time offset $t_0$ and amplitude $A$ are free parameters of the fit.
As an example, 
Figure~\ref{fig:pulseCCE} shows the results of these fits for one of the innermost sensors and for $V_\text{bias}=100$~V and $400$~V. 
The integral of the fitted function between its two zeroes $t_0$ and $t_0+3\tau$ is taken as a measure of the collected charge at the given bias voltage.

\begin{figure}[tb]
\begin{center}
\begin{minipage}[t]{0.49\textwidth}
\begin{overpic}[width=\textwidth,scale=.25,tics=20]{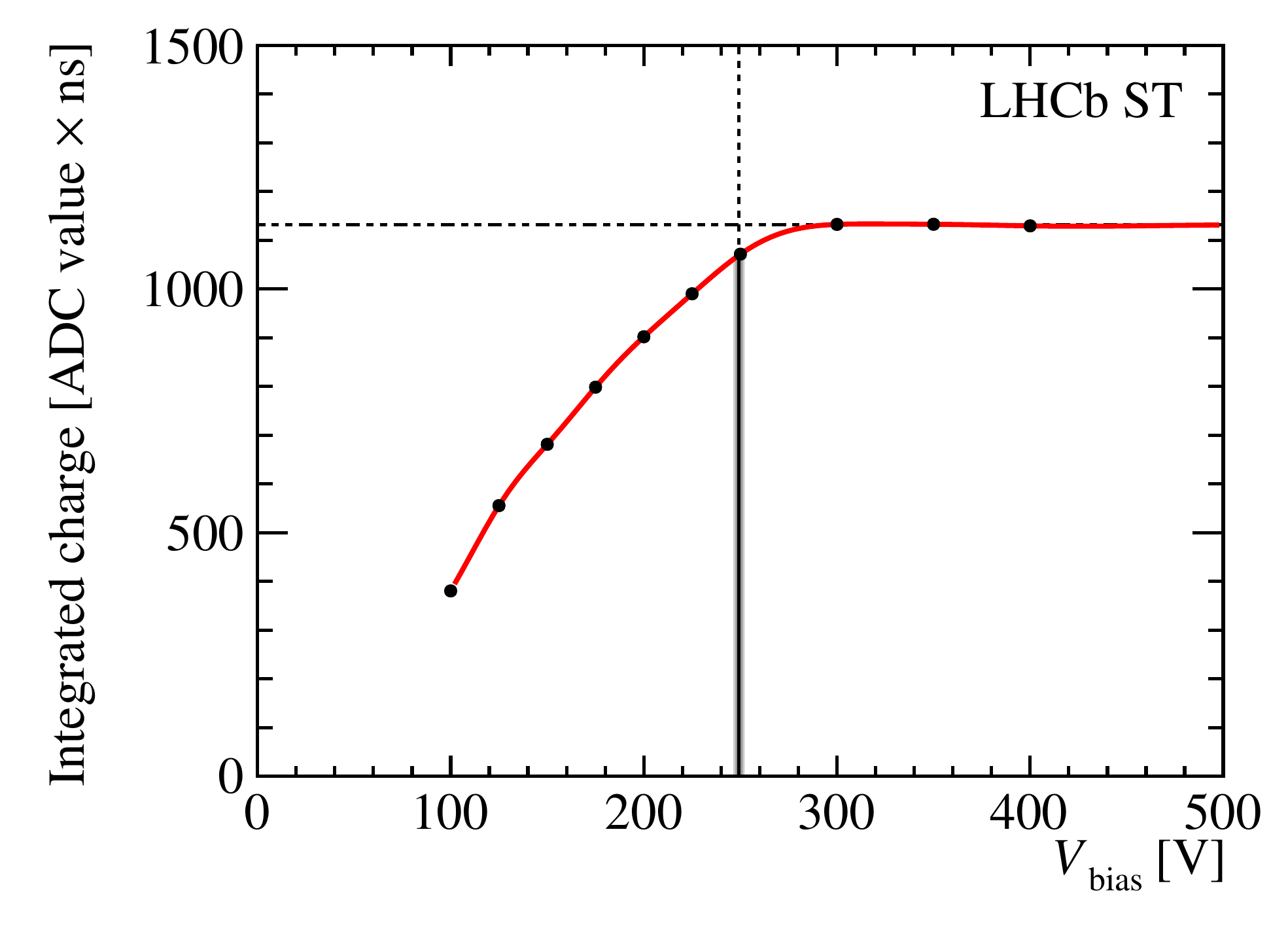}
\put(25,60){(a)}
\end{overpic}
\end{minipage}
\begin{minipage}[t]{0.49\textwidth}
\begin{overpic}[width=\textwidth,scale=.25,tics=20]{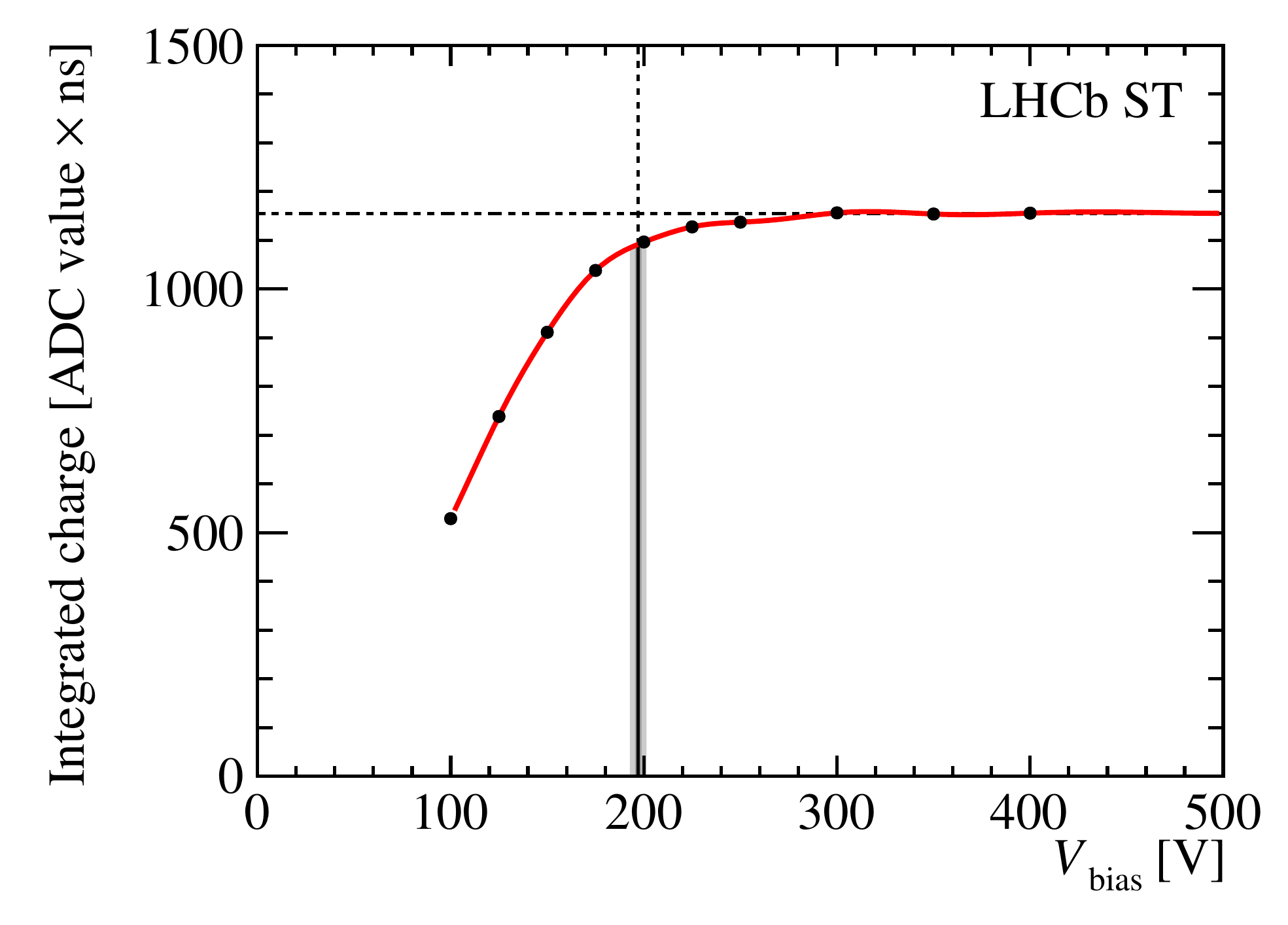}
\put(25,60){(b)}
\end{overpic}
\end{minipage}
\caption[Integrated charge]{Integrated charge as a function of the applied bias voltage for a central TT read-out sector measured in the CCE scans from (a) April 2012 and (b) September 2017. The decrease of $V_\text{depl}$ is clearly visible.}
\label{fig:voltageCCE}
\end{center}
\end{figure}

\begin{figure}[htb]
    \begin{center}
        \begin{minipage}[t]{0.325\textwidth}
            \begin{overpic}[width=\textwidth,scale=.25,tics=20]{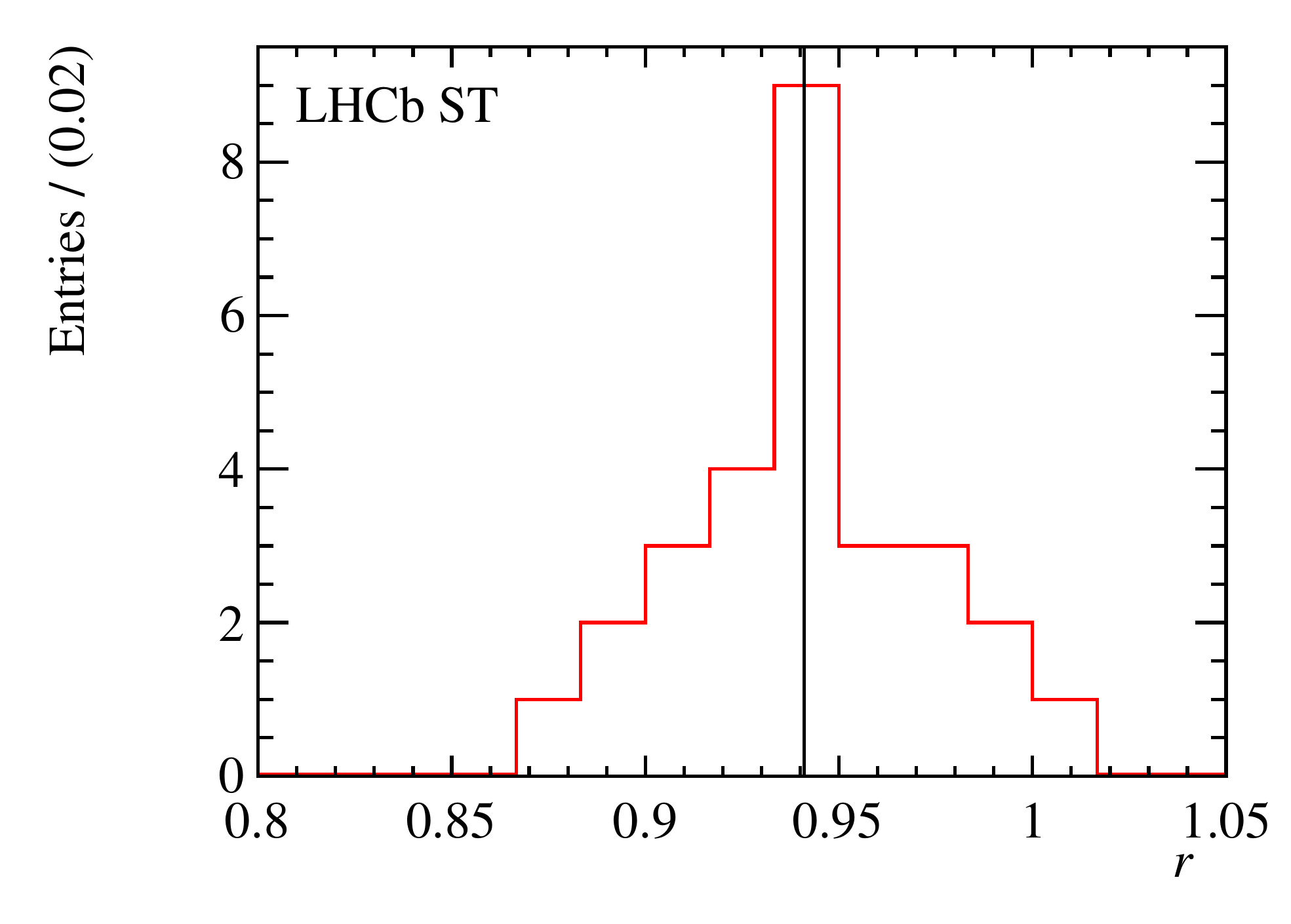}
                \put(79,58){(a)}
            \end{overpic}
        \end{minipage}
        \begin{minipage}[t]{0.325\textwidth}
            \begin{overpic}[width=\textwidth,scale=.25,tics=20]{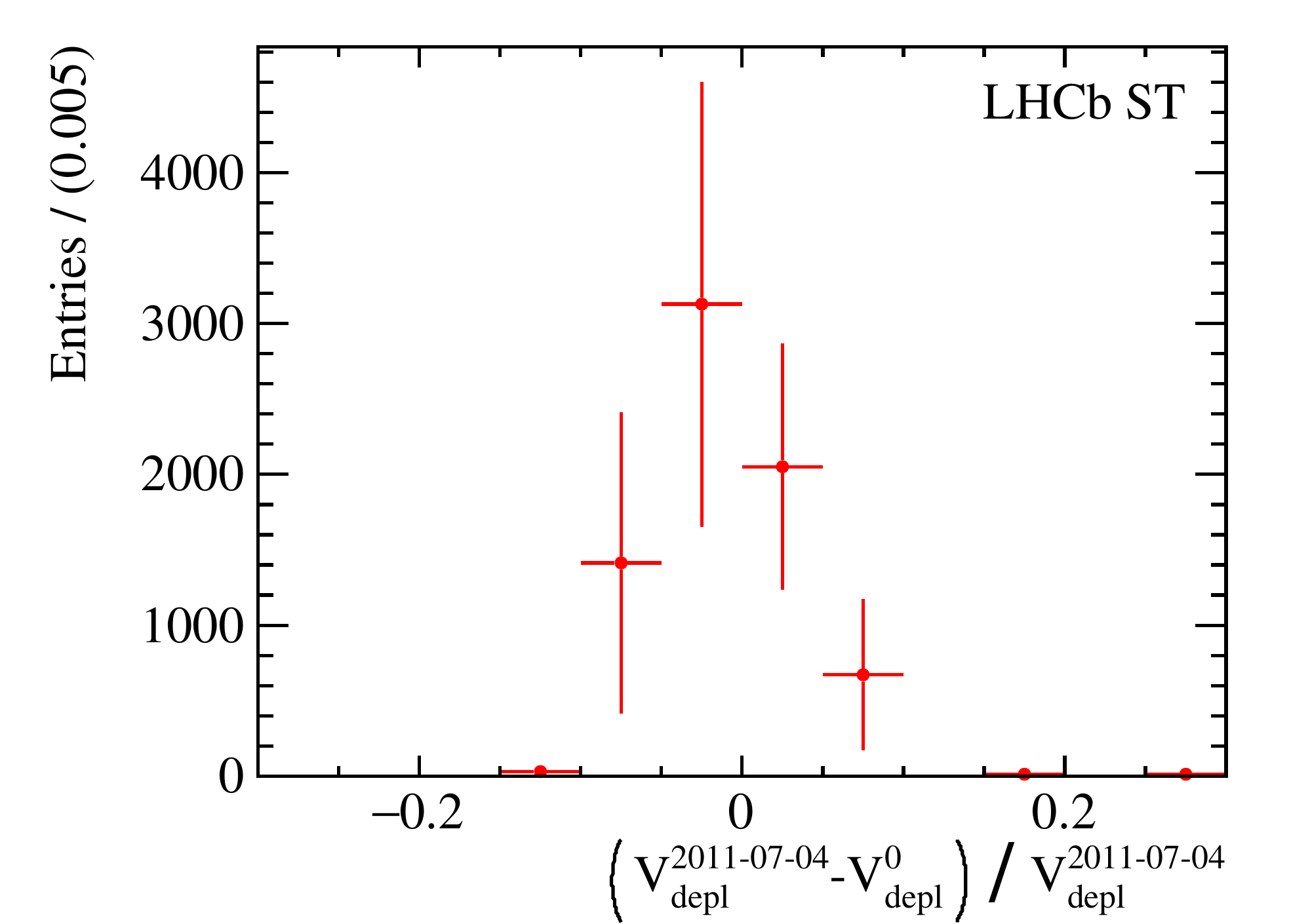}
                \put(25,58){(b)}
            \end{overpic}
        \end{minipage}
        \begin{minipage}[t]{0.325\textwidth}
            \begin{overpic}[width=\textwidth,scale=.25,tics=20]{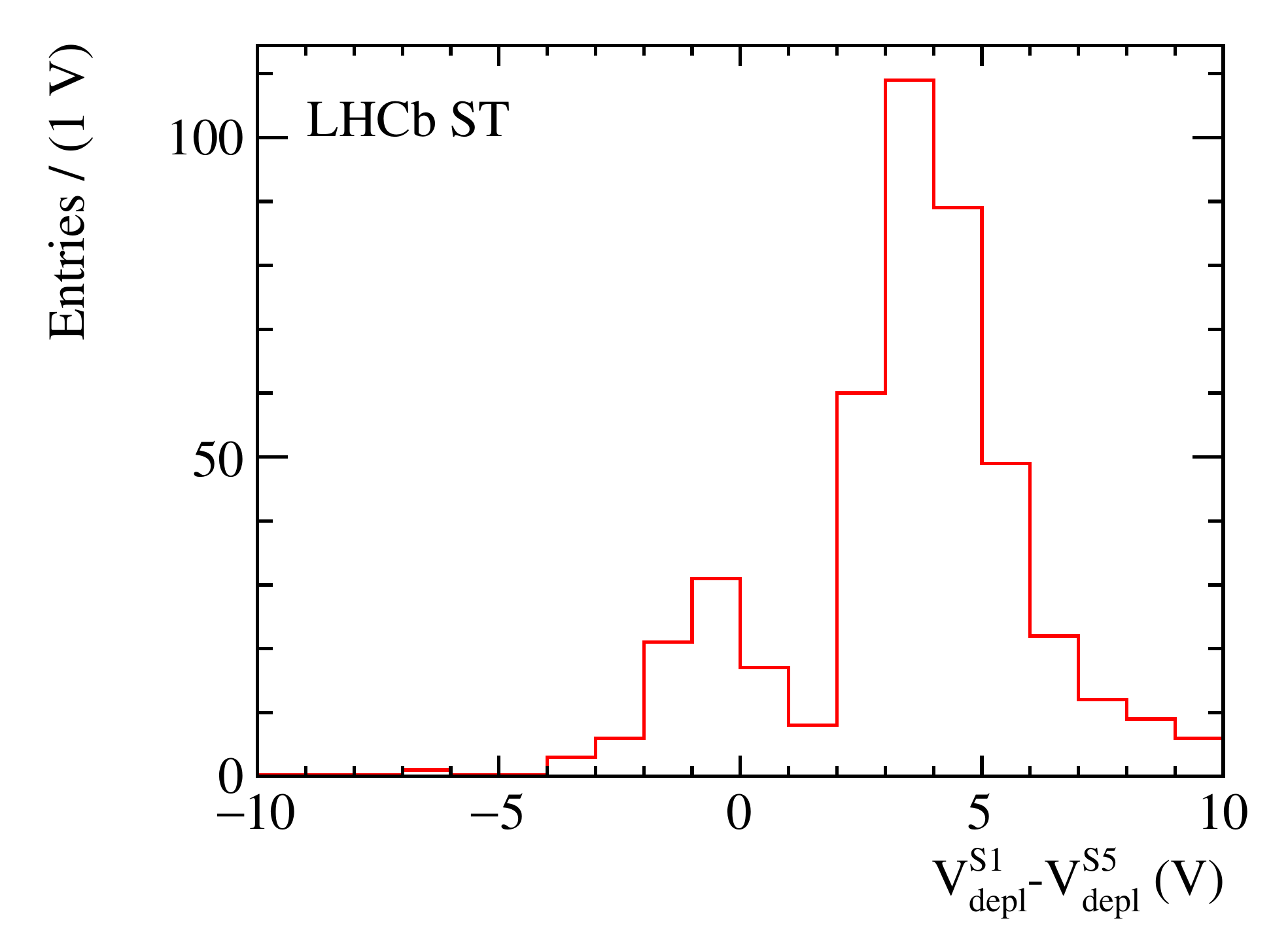}
                \put(79,58){(c)}
            \end{overpic}
        \end{minipage}
        \caption[systematics]{Distributions for all analysed read-out sectors of (a) the ratio $r$ defined in Eq.\ref{eqn:r}; (b) the relative difference between $V_\text{depl}$ and $V_\text{depl}^\text{2011-07-04}$ as discussed in the main text; (c) the differences in $V_\text{depl}$ obtained from a fifth-order spline and from a linear interpolation. The solid black line in (a) highlights the value of the average $\overline{r}$.}
        \label{fig:systs}
    \end{center}
\end{figure}

This quantity, referred to as ``integrated charge'', is shown as a function of $V_\text{bias}$ in Fig.~\ref{fig:voltageCCE}. The three highest $V_\text{bias}$ data points are fit with a constant to determine the plateau value.
The data points are interpolated with a fifth-order spline~\cite{Ferguson:1964:MCI:321217.321225} $S$ and the depletion voltage is determined as the bias voltage at which $S(V_\text{bias})$ reaches the fraction $\overline{r} = 94 \%$ of its value in the plateau.
The value $\overline{r} = 94 \%$ was chosen based on an analysis
of the earliest CCE scan, taken on 2011-07-14. First, the ratio
\begin{align}
r = S\left(V_\text{depl}^\text{2011-07-14}\right)\, /\, S\left(\text{plateau}\right)\label{eqn:r}
\end{align}
was calculated for each analysed read-out sector, where the true depletion voltage at the time of the CCE scan, $V_\text{depl}^\text{2011-07-04}$, was estimated from the full depletion voltage of the sensor as measured in $C-V$ scans during the construction of the detector, corrected for the expected change in full depletion voltage according to the Hamburg model and the integrated fluence received at the time of the CCE scan. 
The distribution of $r$ for all analysed read-out sectors is shown in Fig.~\ref{fig:systs}(a). 
The values of $r$ obtained for the different read-out sectors were then averaged to determine $\overline{r}$.

The systematic uncertainty on $V_\text{depl}$ due to the choice of $\overline{r}$ was estimated by applying the full method to the first CCE scan and comparing the values of $V_\text{depl}$ extracted from the spline fits with $\overline{r} = 94 \%$ to the values $V_\text{depl}^\text{2011-07-04}$ extracted from the initial $C-V$~scans and the Hamburg model. 
The residual distribution of $V_\text{depl}$ and $V_\text{depl}^\text{2011-07-04}$ is shown in Fig.~\ref{fig:systs}(b); the spread of this distribution is taken as systematic uncertainty.

A second source of systematic uncertainty is the choice of a fifth-order spline to interpolate the measurements. 
To estimate this uncertainty, the entire procedure was repeated using a simple linear interpolation between consecutive measurements instead of the spline fit. 
The distribution of the differences in $V_\text{depl}$ obtained using the two types of interpolation are shown in Fig.~\ref{fig:systs}(c) for all analysed read-out sectors; the spread of this distribution is taken as systematic uncertainty.
These two uncertainties are summed in quadrature.

\begin{table}[tb]
    \caption[Parameters for the depletion voltage simulation (Hamburg model)]{Parameters for the depletion voltage simulation by the Hamburg model~\cite{Moll:1999kv}.}
    \begin{center}
        \begin{tabularx}{\textwidth}{p{4.5cm}r@{}l@{}l}
            \toprule
            \textbf{Parameter} & \multicolumn{3}{c}{\textbf{Value}}\\
            \midrule
            $n_{c,0}$ & $(3.28\pm0.26)$ & $\times 10^{-10}$ & cm$^{-3}$ \\
            $c$       & $2.29\color{white})$ & $\times 10^{-13}$ & cm$^2$\\
            $g_c$     & $(1.60\pm0.04)$ & $\times 10^{-2}$ & cm$^{-1}$\\
            $g_a$& $(1.40\pm0.14)$ & $\times 10^{-2}$ & cm$^{-1}$\\
            $g_r$& $(5.70\pm0.09)$ & $\times 10^{-2}$ & cm$^{-1}$\\
            $k_{a,0}$& $(2.4\enspace\pm1.0\enspace)$ & $\times 10^{15}$ & s$^{-1}$\\
            $k_{r,0}$& $(1.5\enspace\pm1.1\enspace)$ & $\times 10^{15}$ & s$^{-1}$\\
            $E_{aa}$ & $(1.09\pm0.03)$ &  & eV\\
            $E_{ar}$ & $(1.31\pm0.03)$ &  & eV\\
            \bottomrule
        \end{tabularx}
    \end{center}
    \label{tab:HamburgPara}
\end{table}

The change $\Delta\neff$ in the effective doping concentration is described in the Hamburg model by three different mechanisms, described in detail in Refs.~\cite{Wunstorf:1992ua,Fretwurst:1994nq,Feick:1997ec,Moll:1999kv}:
\begin{enumerate}
	\item a contribution associated to the effective (incomplete) removal of donor atoms and the addition of stable acceptor atoms due to radiation-induced changes in the band structure of the silicon
crystal. This contribution, referred to as ``stable damage'', is described by
	\begin{equation}
	\Delta n_c(\feq) = n_{c,0}\,\big[1-\exp(-c\,\feq)\big] + g_c\,\feq;
	\label{eq:HamburgStableDamage} 
	\end{equation}
	\item a temperature-dependent contribution due to the annealing of the induced defects, which can be described by
	\begin{equation}
	\Delta n_a(\feq,\,t,\,T_a) = \feq \, g_a\exp\bigg(-\frac{t}{\tau_a(T_a)}\bigg);
	\label{eq:HamburgAnnealing} 
	\end{equation}
	\item a contribution due to the combination of individual defects in the silicon lattice, leading to stable defects (``reverse annealing''), described by
	\begin{equation}
	\Delta n_r(\feq,\,t,\,T_a) = \feq \, g_r\bigg(1-\frac{1}{1+t/\tau_r(T_a)}\bigg).
	\label{eq:HamburgReverseAnnealing} 
	\end{equation}
\end{enumerate}
The sum of the three contributions describes the total change of $\neff$: 
\begin{equation}
    \Delta\neff = \Delta n_c + \Delta n_a + \Delta n_r.
\end{equation}
The values of the model parameters are listed in Table~\ref{tab:HamburgPara}. Their values were measured in dedicated irradiation campaigns and can be found in Reference~\cite{Moll:1999kv}.
The parameters $\tau_a$ and $\tau_r$ have a temperature dependence described by the Arrhenius relation, \cf Eq.~\ref{eq:Arrhenius}, with parameters $E_{aa}$ and $k_{a,0}$ and $E_{ar}$ and $k_{r,0}$, respectively, and the ambient temperature $T_a$ is taken from sensors placed inside the detector boxes.

\begin{figure}[tb]
	\begin{center}
		\begin{minipage}[t]{0.49\textwidth}
			\begin{overpic}[width=\textwidth,scale=.25,tics=20]{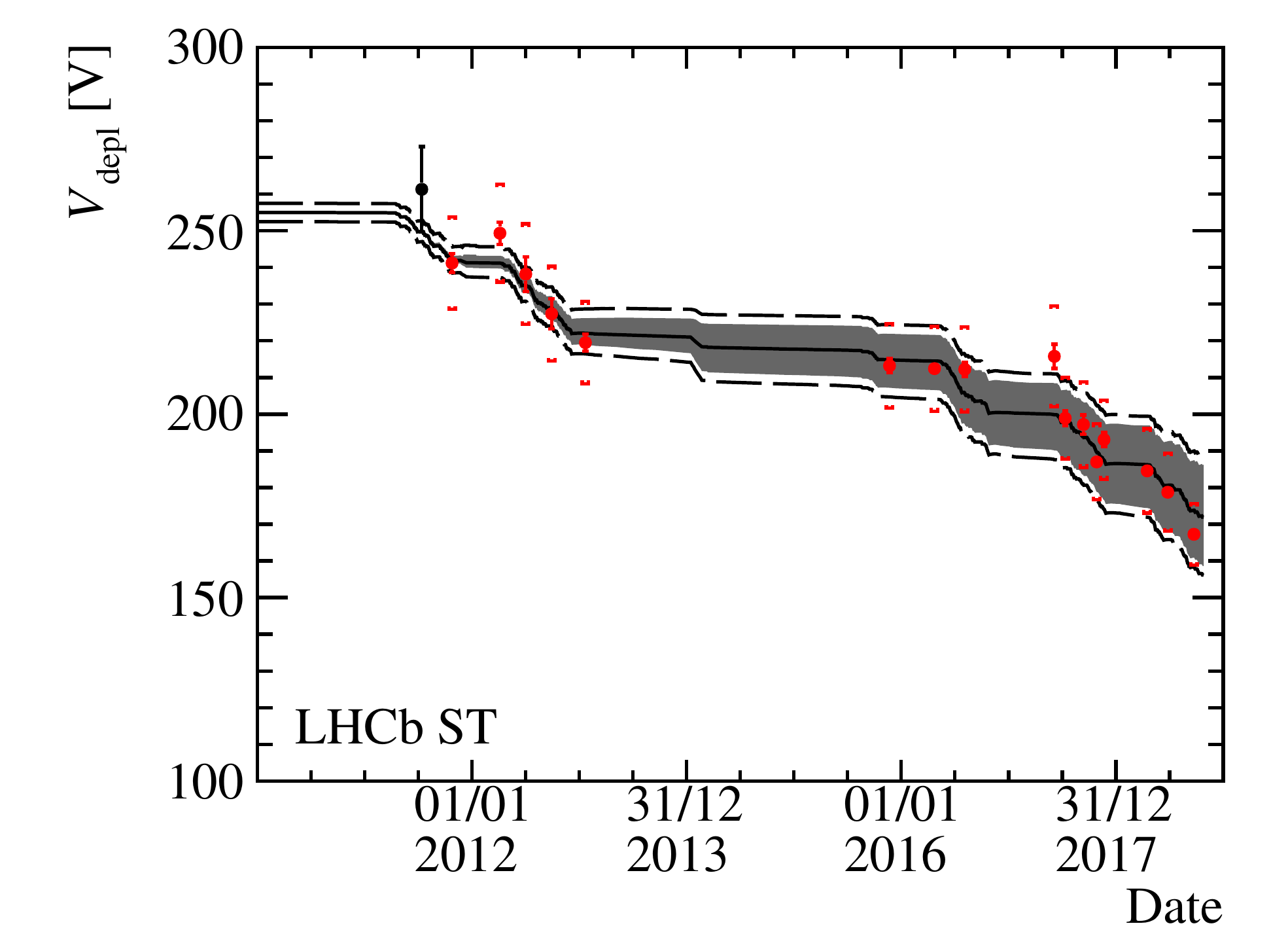}
				\put(80,60){(a)}
			\end{overpic}
		\end{minipage}
		\begin{minipage}[t]{0.49\textwidth}
			\begin{overpic}[width=\textwidth,scale=.25,tics=20]{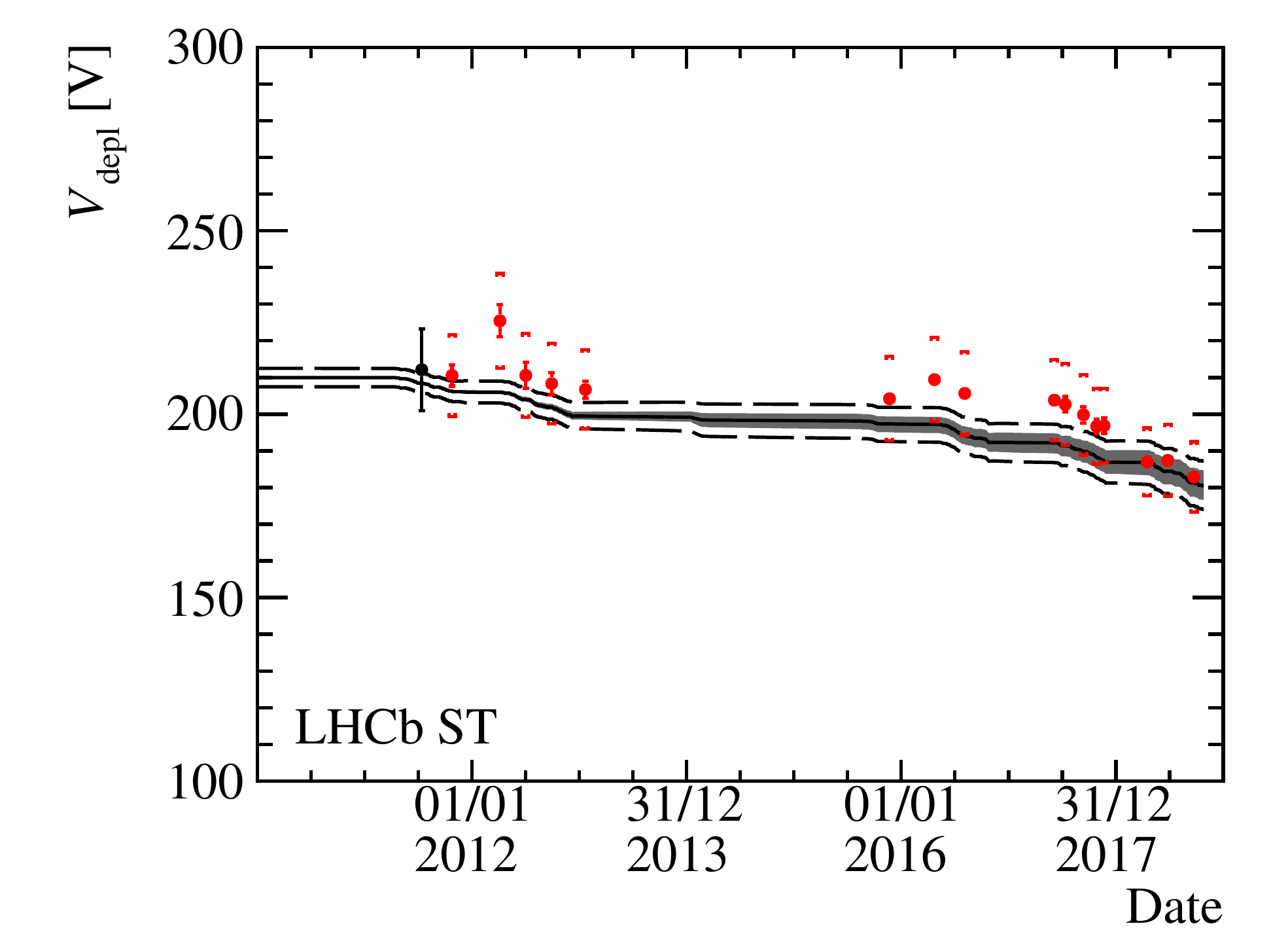}
				\put(80,60){(b)}
			\end{overpic}
		\end{minipage}
		\caption[$V_\text{depl}$ of single sensors]{Measured effective depletion voltage (red points) in the area of two TT sensors: one just above the beam pipe (a) and one further away (b), whose positions are indicated in Figure~\ref{fig:sketches}(a). The black point and error bars correspond to the CCE scan used to calibrate the ratio $r$. The statistical contribution to the total uncertainty is indicated by the solid error bars. The predicted evolution of the depletion voltage, based on the initial depletion voltage measured after sensor production, the running conditions and the Hamburg model, is shown as a solid black line. The grey bands show the uncertainty on the predicted evolution of $V_\text{depl}$, while the black dashed lines account for the $\pm 2.5$~V uncertainty on the measurement of  the initial depletion voltage $V^0_\text{depl}$ in $C-V$ scans.}
		\label{fig:singleSensor}
	\end{center}
\end{figure}

\begin{figure}[tbp]
    \begin{center}
        \begin{minipage}[t]{0.4\textwidth}
            \begin{overpic}[width=\textwidth,scale=.25,tics=20]{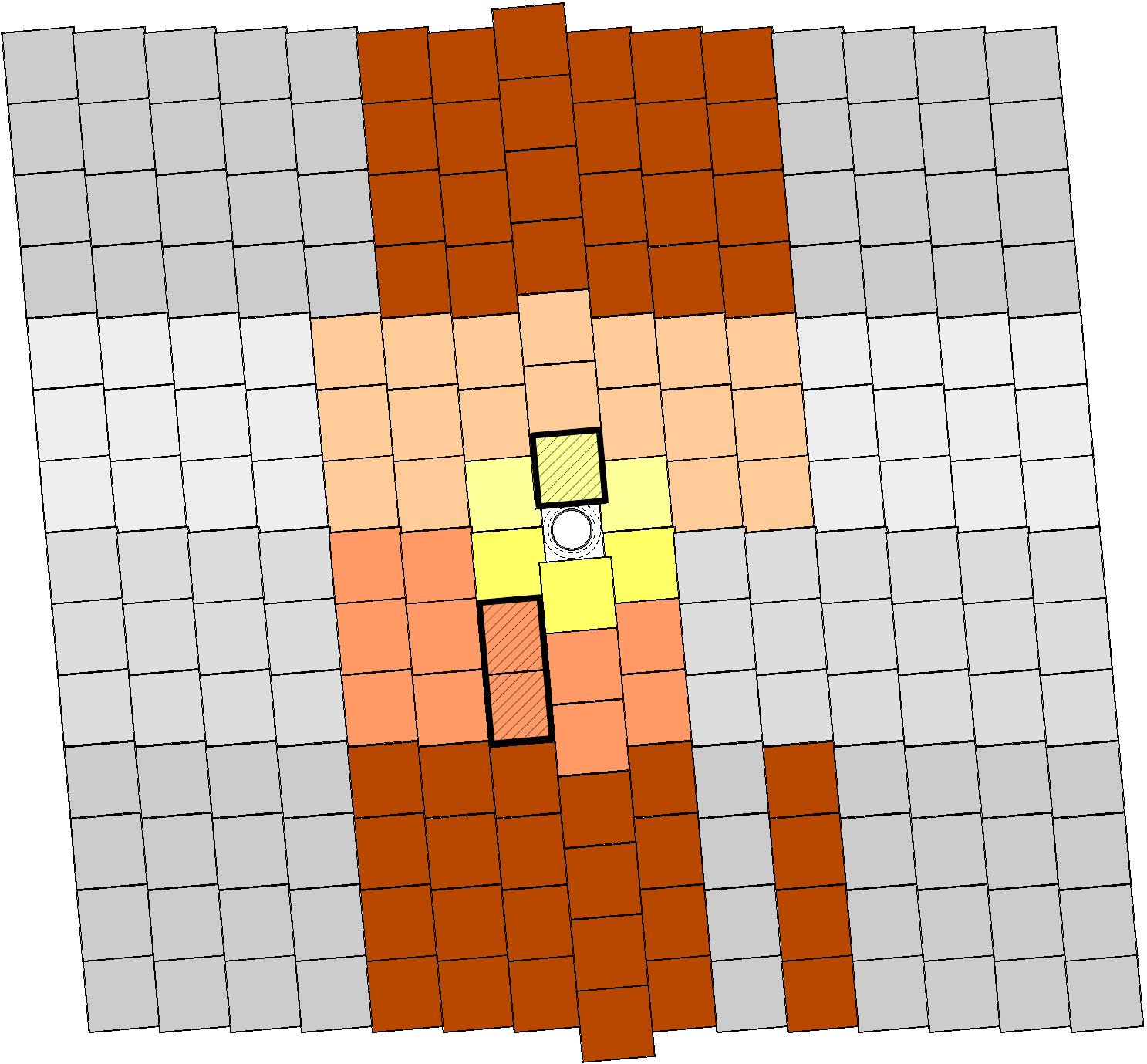}
                \put(42,-7){(a)}
            \end{overpic}
        \end{minipage}\\[25pt]
        \begin{minipage}[t]{0.49\textwidth}
            \begin{overpic}[width=1.05\textwidth,scale=.25,tics=20]{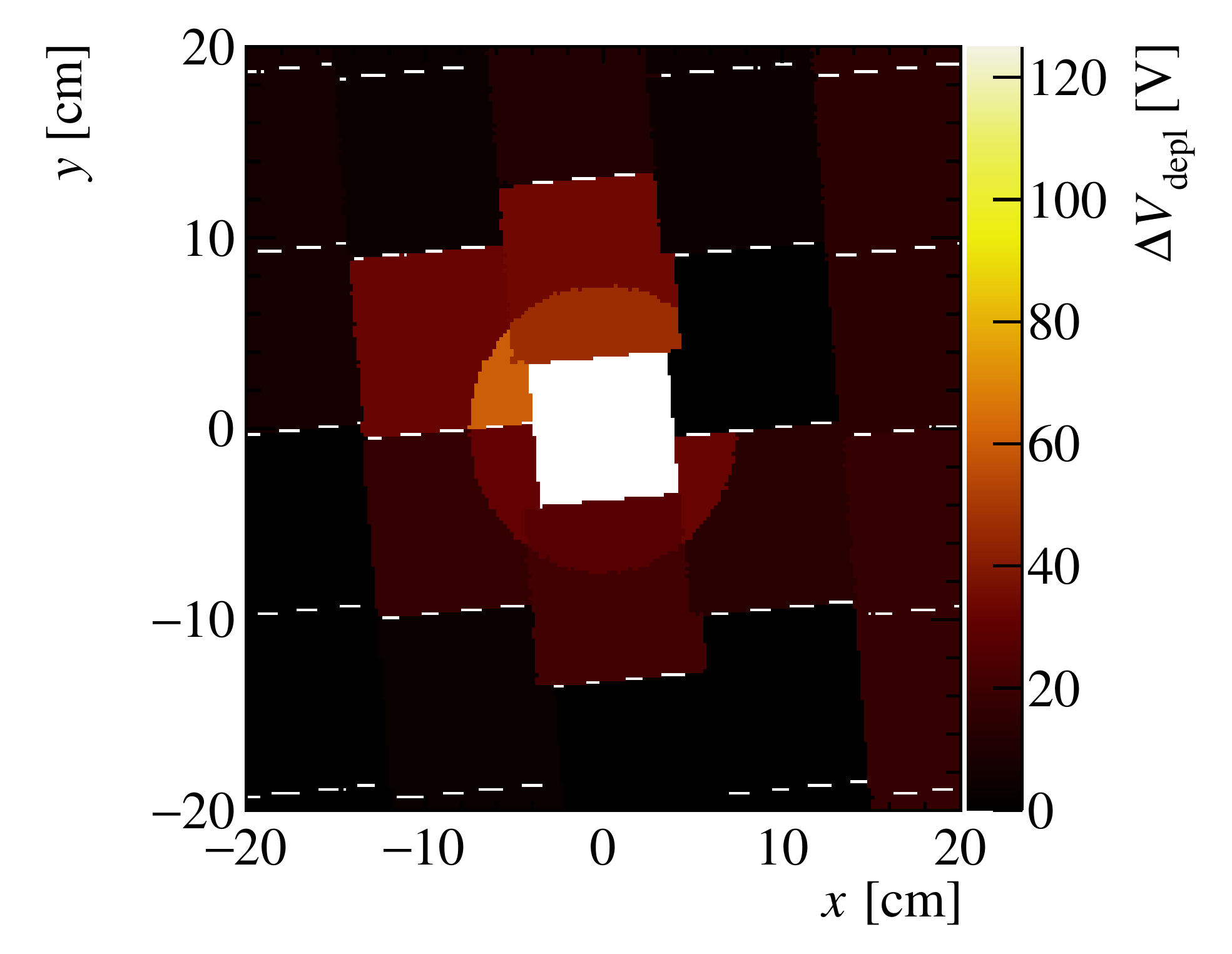}
                \put(40,0){(b)}
            \end{overpic}
        \end{minipage}
        \begin{minipage}[t]{0.49\textwidth}
            \begin{overpic}[width=1.05\textwidth,scale=.25,tics=20]{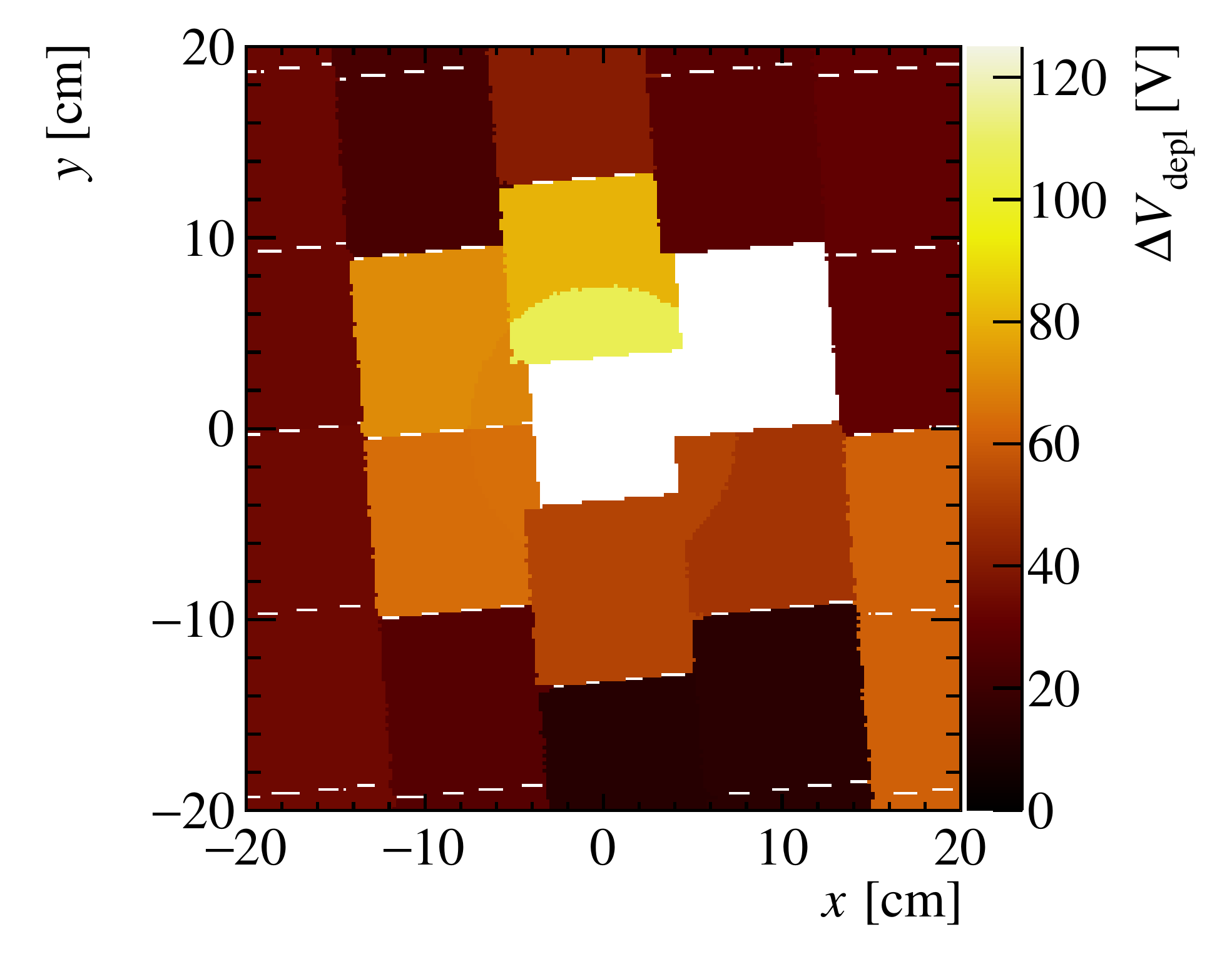}
                \put(40,0){(c)}
            \end{overpic}
        \end{minipage}
        \caption[$V_\text{depl}$ of single sensors]{Sketch highlighting the TT sectors analysed during the various CCE scans (a). The greyed out sectors were excluded due to poor statistics. The two sectors highlighted with dashes, just above the beam pipe and below the beam pipe on the left hand side, are those which evolution is shown in Figure~\ref{fig:singleSensor}(a) and Figure~\ref{fig:singleSensor}(b), respectively. Absolute change in $V_\text{depl}$ in the innermost region of TT in  January 2013 and September 2018  is shown in (b) and (c) respectively. The white sector in (c) became inoperative in August 2017 due to an HV connector issue.}
        \label{fig:sketches}
    \end{center}
\end{figure}

Figure~\ref{fig:singleSensor} shows the measured depletion voltages for a one-sensor and for a two-sensors TT read-out sectors, highlighted in Figure~\ref{fig:sketches}(a), and their predicted evolution based on the Hamburg model described above. 
This calculation uses the fluence estimated from the \textsc{Fluka} simulations\footnote{Fluence is calculated by integrating the radiation map over the sensor area.}, the actual running conditions and the temperature measurements in the detector boxes.

\begin{figure}[htb]
    \begin{center}
        \begin{minipage}[t]{0.49\textwidth}
            \begin{overpic}[width=\textwidth,scale=.25,tics=20]{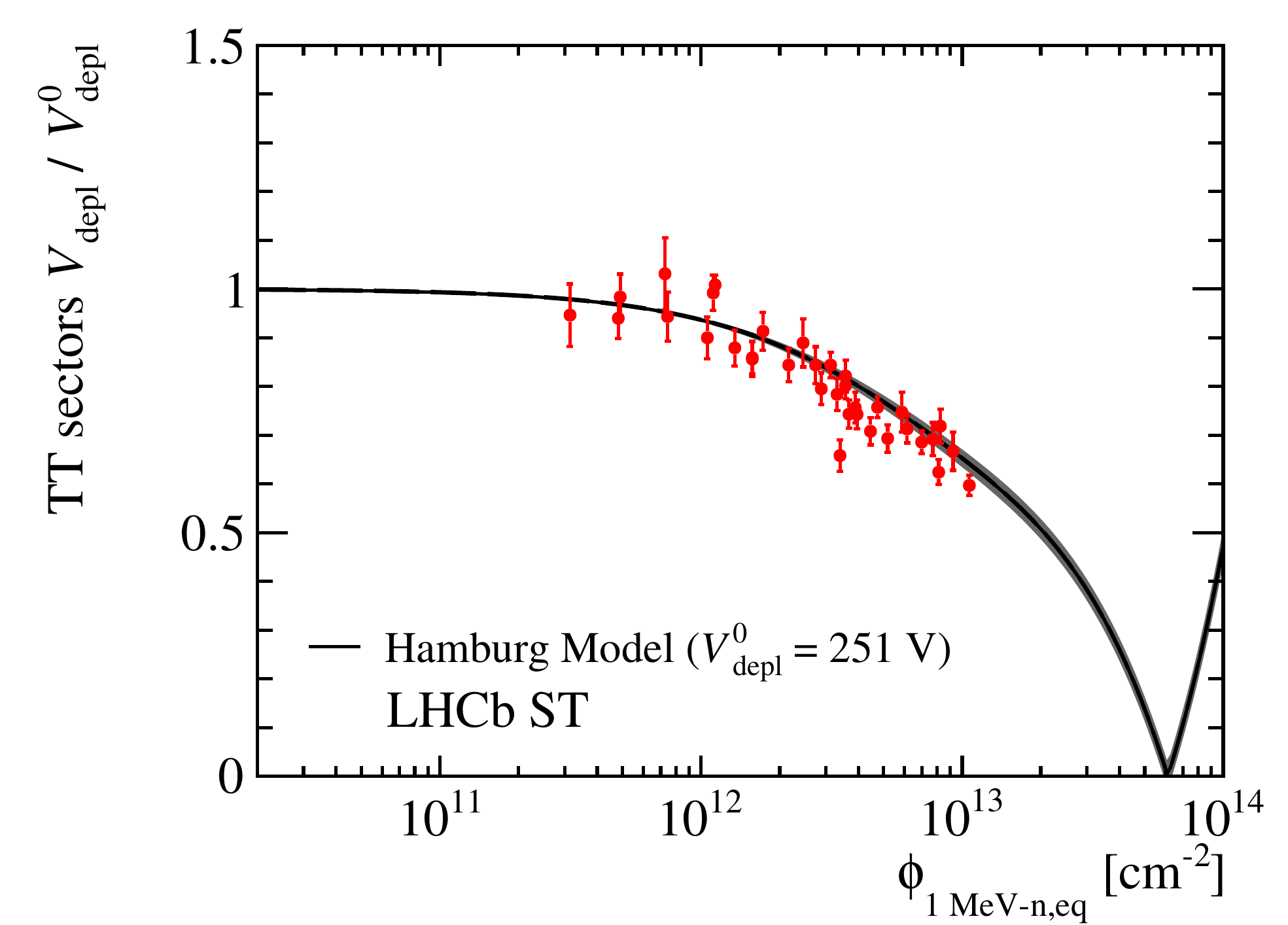}
                \put(80,60){(a)}
            \end{overpic}
        \end{minipage}
        \begin{minipage}[t]{0.49\textwidth}
            \begin{overpic}[width=\textwidth,scale=.25,tics=20]{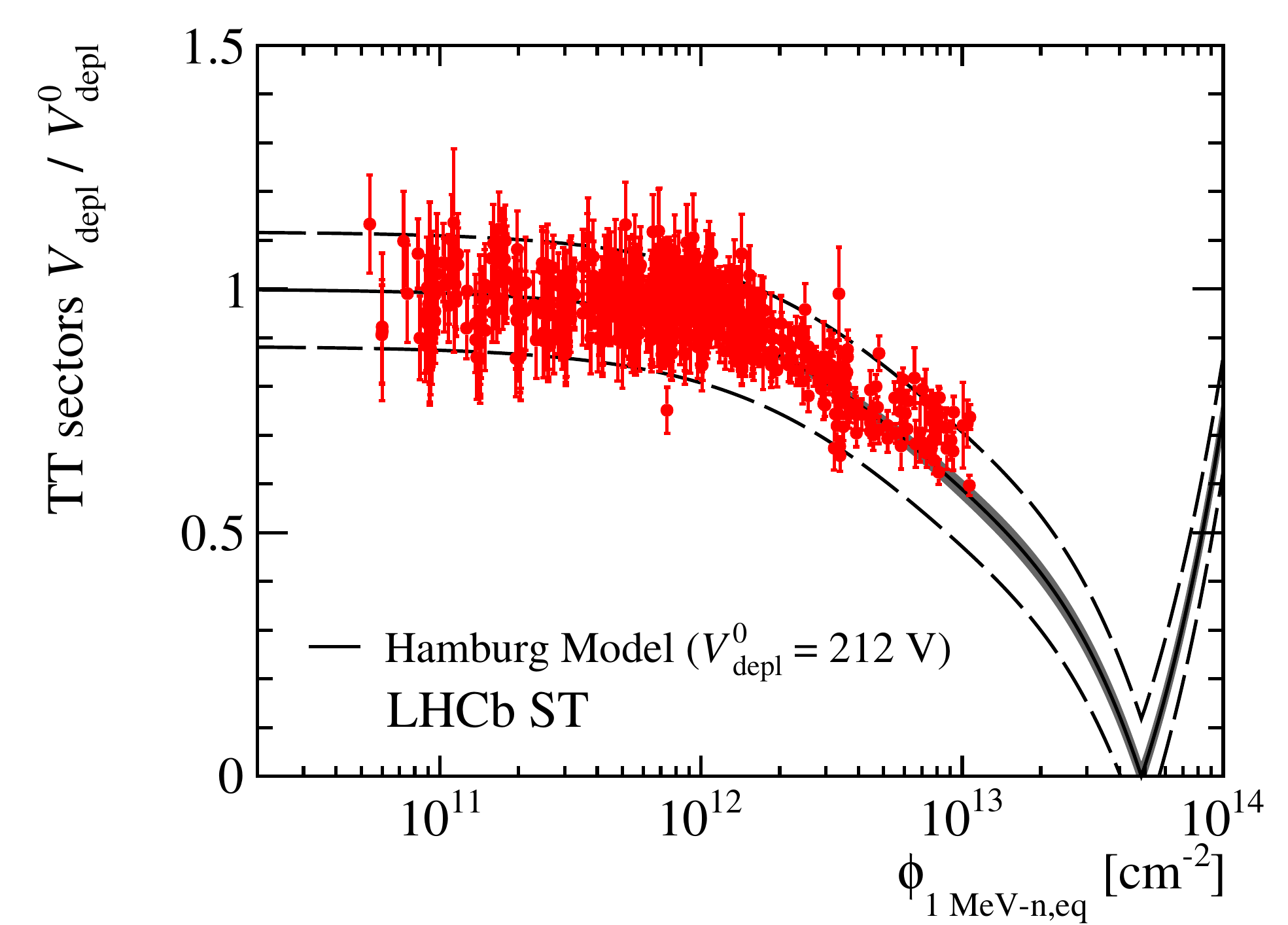}
                \put(80,60){(b)}
            \end{overpic}
        \end{minipage}
\caption[Evolution of $V_\text{depl}$ versus $\feq$]{Measured evolution of $V_\text{depl}$ as a function of the 1-MeV neutron equivalent fluence obtained from the running conditions and \textsc{Fluka} for a TT sector in the area close to the beam pipe~(a) and for all of the read-out sectors in the TT~(b). The innermost sectors are subdivided in regions separating tracks traversing the detector at a distance larger or smaller than 75~mm from the beam axis. The error bars of the data points display the sum of the statistical and systematic uncertainty. The solid black curve shows the predictions based on the stable damage part of the Hamburg model, the grey shaded region its uncertainty due to the parameter uncertainty of the model. The initial depletion voltage $V^0_{\text{depl}}$ for the Hamburg model prediction in~(b) is averaged among all sectors, and the dashed black lines show the standard deviation of the distribution of the initial $V^0_\text{depl}$ values.}
\label{fig:deplVphiN}
\end{center}
\end{figure}

For the read-out sectors closest to the LHC beam pipe, dedicated measurements only including tracks traversing the sensors within 75~mm from the beam axis are performed.
Fig.~\ref{fig:deplVphiN}~shows the measured values of $V_\text{depl}$ of the different TT read-out sectors in the different CCE scans as a function of the 1-\mev-neutron equivalent fluence estimated from the \textsc{Fluka} simulations and the integrated luminosity collected in LHCb. 
It also shows the expected evolution of $V_\text{depl}$ based on the stable damage contribution $n_c$.

Fig.~\ref{fig:sketches}(c) shows the absolute change of $V_\text{depl}$ in the innermost part of the detector between July 2011 and September 2018.

Good agreement between the measurements and the predicted evolution of $V_\text{depl}$ is observed, with no type inversion seen in the silicon sensors of the TT at the end of the LHC Run~2.

\section{Conclusion\label{sec:summary}}

The evolution of the radiation damage in the LHCb Tracker Turicensis has been monitored using measurements of leakage currents and effective depletion voltages. 
The latter were performed with data collected in dedicated charge collection efficiency scans. 
The obtained results show good agreement with predictions based on phenomenological models. 
At the end of 2018, the innermost sensors, which experience the highest fluence, had not yet reached the point of type inversion, and no modifications to the operation procedure of the detector were performed in its last year of operation.
The detector has been dismantled and will be replaced as part of the LHCb upgrade~\cite{LHCb-TDR-012,*LHCb-TDR-015} during the long shutdown LS2 of the LHC.

\section*{Acknowledgements}
E.~G. gratefully acknowledges support by the Forschungskredit of the University of Zurich, grant no.~FK-17-108.
The work of M.~A. is supported by the Swiss National Science Foundation (SNF) under contract BSSGI0\_155990.

\clearpage
\addcontentsline{toc}{section}{References}
\setboolean{inbibliography}{true}
\bibliographystyle{bib/LHCb}
\bibliography{bib/raddam,bib/LHCb-DP,bib/LHCb-TDR}

\end{document}